%% file: mass-malt90-contreras_astroph.tex
\pdfoutput=1

\documentclass[useAMS,usenatbib]{mn2e}

\usepackage{subfigure}
\usepackage{longtable,lscape}
\usepackage{threeparttable}
\usepackage{verbatim}
\usepackage{rotating}
\usepackage{color}
\usepackage{amsmath}

\newcommand{\comments}[1]{}

\newcommand\nhp{\mbox{N$_2$H$^+$}}

\newcommand\hnc{HNC}

\newcommand\Msun{M$_\odot$}

\newcommand{\mum}{$\mu$m}

\bibpunct{(}{)}{;}{a}{}{,}

\defcitealias{Contreras-2013b}{Contreras et al. 2013b} 

\def\nmalt{3,557}	

\def\ymca{AGAL331.029-00.431\_S}
\def\ymcb{AGAL331.034-00.419\_S}

\newcommand{\ncoherent}{44}
\newcommand{\nmaltused}{1,244}
\newcommand{\ncoldclumps}{24}

\title[Cluster precursors in the MALT90 survey]{Characterizing the properties of cluster precursors in the MALT90 survey}

\author[Yanett Contreras et al.]{Yanett Contreras$^{1,2}$\thanks{E-mail: ycontreras@strw.leidenuniv.nl},  Jill M. Rathborne$^{2}$,
Andres Guzman$^{3}$, James Jackson$^{4}$,
\newauthor Scott Whitaker$^{5}$, Patricio Sanhueza$^{6}$ and Jonathan Foster$^{7}$\\
   $^{1}$Leiden Observatory, Leiden University, PO Box 9513, NL-2300 RA Leiden, the Netherlands\\
    $^{2}$CSIRO Astronomy and Space Science, P.O. Box 76, Epping NSW 1710, Australia\\
  $^{3}$Departamento de Astronom\'ia, Universidad de Chile, Camino el Observatorio 1515, Las Condes, Santiago, Chile\\
  $^{4}$School of Mathematical and Physical Sciences, University of Newcastle, University Drive, Callaghan NSW 2308, Australia\\
  $^{5}$Physics Department, Boston University, Boston, MA, USA\\
  $^{6}$National Astronomical Observatory of Japan, National Institute of Natural Sciences, 2-21-1 Osawa, Mitaka, Tokyo 181-8588, Japan\\
  $^{7}$Department of Astronomy, Yale University, P.O. Box 28101 New Haven, CT 06520-8101, USA\\}
\begin{document}

\date{Accepted 2016 November 26. Received 2016 November 26; in original form 2016 September 2}

\maketitle

\label{firstpage}

\begin{abstract}
In the Milky Way there are thousands of stellar clusters each harboring from a hundred to a million  stars. Although clusters are common, the initial conditions of cluster formation are still not well understood.  To determine the processes involved in the formation and evolution of clusters it is key to determine the global properties of cluster-forming clumps in their earliest stages of evolution. Here, we present the physical properties of \nmaltused~clumps identified from the MALT90 survey. Using the dust temperature of the clumps as a proxy for evolution we determined how the clump properties change at different evolutionary stages. We find that less-evolved clumps exhibiting dust temperatures lower than 20 K have higher densities and are more gravitationally bound than more-evolved clumps with higher dust temperatures. We also identified a sample of clumps in a very early stage of evolution, thus potential candidates for high-mass star-forming clumps. Only one clump in our sample has physical properties consistent with a young massive cluster progenitor, reinforcing the fact that massive proto-clusters are very rare in the Galaxy.

\end{abstract}

\begin{keywords}
ISM: clouds -- stars: formation -- submillimetre: ISM -- surveys
\end{keywords}

\section{Introduction}

Stars form within dense ($>10^4$ cm$^{-3}$) small cores ($<$0.1 pc). Most stars are not born in isolation: these small cores are usually embedded in larger structures commonly referred as clumps ($\sim$1-2 pc) that have a wide range of masses, sizes, and are located in different environments. These clumps will evolve to harbor stars and to form complex stellar systems, or clusters. In the Milky Way there are thousands of open clusters, each harboring from 100 to 10$^4$ stars, and hundreds of globular clusters, each harboring from 10$^4$ to over 10$^6$ stars \citep[e.g.,][]{Bressert-2012}.  However, although clusters are common in the Galaxy, the initial conditions of cluster-formation are still not well understood. 

To understand how clusters are formed it is necessary to identify and characterize the physical properties of cluster-forming clumps in a range of evolutionary stages: from starless clumps to those harboring deeply embedded young stellar objects.

Due to the relative proximity of low-mass star-forming regions, clumps giving rise to low-mass stars have been systematically better characterized than their intermediate- and high-mass counterparts \citep[e.g.][]{Tafalla-2004, Andre-2007}. High-mass star-forming clumps, however, are usually located at greater distances and embedded in complex environments which have made their study more challenging. To overcome these difficulties, recent unbiased Galactic plane surveys have been key to determining the physical properties of cluster-forming clumps and to identifying proto-cluster candidates \citep[e.g.][]{Ginsburg-2012, jackson-2013, Hoq-2013, Urquhart-2014, Guzman-2015, Svoboda-2016}. 

To determine if clumps can evolve into clusters harboring high-mass stars it is necessary to find the physical properties of the cluster-forming clump progenitors. \citet{Kauffmann-2010b} used observations of local molecular clouds to determine a empirical threshold for clouds harbouring high-mass stars. They showed that these clouds follow an empirical relationship between their size and mass, defining a threshold for high-mass star-forming clouds given by $M(r)\geq 870 M_\odot (r/\textrm{pc})^{1.33}$, with $r$ the radius of the cloud.

In a similar manner, one can determine the potential of the clumps to evolve into open or young massive clusters (YMCs). \citet{Portegies-2010} determined the global properties of open clusters in the Galaxy, providing constraints to the physical properties that open cluster progenitors ought to have. YMCs, with masses of $\geq10^4$ M$_\odot$, might represent the missing link between open (M$<10^4~{\rm M}_\odot$) and globular clusters  (M $\sim10^4-10^6~{\rm M}_\odot$). Therefore, the study of the physical properties of the YMC progenitors may provide key information about the formation mechanisms of globular clusters, and might contribute to our understanding of cluster formation in the early universe \citep{Longmore-2014}. Clumps forming YMCs ought to have masses $\geq 3\times10^4$ M$_\odot$ (assuming a star-forming efficiency of 30\%).  Their gravitational escape velocity needs to be greater than the photo-ionized gas sound speed so that ionization of newly formed stars will not disrupt the natal gas and the cluster will remain bound. Since observations of YMCs suggest that they are in -- or close -- to virial equilibrium at ages of $\sim$1 Gyr, one can assume that YMCs must have gone through at least through one crossing time of 1 Gyr. These constraints provide a parameter space for the expected physical properties for YMC progenitors: a mass of $\rm{M_{clump}}> 3\times10^4~\rm{M}_\odot$, and a size of $r > r_\Omega = 2G\rm{M}_{clump}c_s^{-2}$ for $\rm{M}_{clump} < 8.4\times10^4\rm{M}_\odot$, and $r > r_{vir} \sim (G{\rm M_{clump}}t^2_{cross})^{1/3}$ with $t_{cross}\sim1$ Gyr, for $\rm{M}_{clump} > 8.4\times10^4\rm{M}_\odot$ \citep{Bressert-2012}.
 
YMC progenitors seem to be rare in the Galaxy. \citet{Ginsburg-2012} used BOLOCAM dust continuum data to search for YMC progenitors in the Galactic plane. They found 3 clumps in the solar neighborhood with characteristics of YMCs progenitors. However, all these clumps showed evidence for active star formation within them.  The only YMC progenitor candidate in an early stage of evolution that has been found so far is G0.253$+$0.016, or ``The Brick". ``The Brick" \citep{Rathborne-2014a, Rathborne-2015}, is an extremely dense ($>10^4~{\rm cm}^{-3}$), massive ($\sim10^5~{\rm M}_\odot$) and cold ($\sim20$ K) molecular cloud \citep{Lis-1994,Lis-1998,Lis-2001} located in the harsh environment of the Galactic Centre. The environmental conditions in the Galactic Centre are very different from those  of the solar neighborhood.  The column density, gas temperature, velocity dispersion, interstellar radiation field, pressure and cosmic ray ionization rate range from being factors of a few to several orders of magnitude higher compared to the solar neighborhood \citep{Walmsley-1986,Morris-1996,Ao-2013}. The difficulty of finding YMCs progenitors outside the Galactic Center might also be due to their fast evolution. Indeed \citet{Svoboda-2016} estimated the life time of YMC progenitors to be 0.03 Myr. Thus, to determine whether starless YMC progenitors exist outside the Galactic Center we need to extend this search to large samples of molecular clumps in other parts of the Galaxy.

In this paper we present the global properties of \nmaltused~clumps
observed in the MALT90 survey. Their wide range of evolutionary stages allows us to analyze their physical
properties as a function of their evolution. \S 2 describes the
data and the physical properties of the clumps: volume densities, masses, virial masses, and bolometric luminosities. In \S 3 we
discuss the differences in the physical properties and virial states of the
clumps as a function of their dust temperature, used to indicate their evolutionary stage. We identify clumps that could evolve into clusters harboring high-mass stars, open clusters and
YMCs, and we determine the properties of cluster-forming clumps in their earliest evolutionary stage. Finally, \S 4 summarizes the main
results of this paper.

\section{cluster-forming clumps within the Galaxy}

Multi-wavelength Galactic plane surveys are extremely valuable for identifying and 
characterizing, in a systematic way, the properties of cluster-forming clumps within 
the Galaxy. Infrared surveys such as the \textit{Spitzer} 3--8 $\mu$m GLIMPSE \citep{benjamin-2003} 
and 24 \mum ~MIPSGAL  \citep{carey-2005} surveys can be used to pinpoint regions of active star 
formation. In particular, the 4.5 \mum~\textit{Spitzer} band is commonly used as a tracer of shocked gas, as the enhancement of the emission at this wavelength is associated with the excitation of shocked H$_2$ ($v = 0-0$) S(9,10,11) lines and/or CO ($v = 1-0$) band-heads. Emission from embedded proto-stars is usually detected by their emission at 8 \mum~and at 24 \mum~due primarily to thermal dust emission \citep{Churchwell-2009}. Once the young stars start to ionize their environment, they will form a compact region of ionized hydrogen (HII region) that can be also detected at 8 \mum~due primarily to bright polycyclic aromatic hydrocarbon (PAH) emission and at 24 \mum~due to their thermal dust emission.

Galactic plane surveys  of the dust continuum emission (e.g. ATLASGAL, \citet{schuller}; 
BGPS, \citet{bolocam}; Hi-Gal, \citet{hershel}) provide a complete census of the dust 
emission toward cluster-forming clumps across the Galaxy \citep{schuller}.  Follow-up surveys of 
their dense molecular gas \citep[e.g., MALT90;][]{foster-2011,jackson-2013, Rathborne-2016} have been also key since these data can be used to 
characterize the gas chemistry and kinematics as well as to provide information about the clumps' kinematic distances.
Moreover, these data are important for determining whether or not adjacent dust continuum peaks 
are in fact coherent, physically related structures \citep[e.g.][]{Contreras-2013b, Contreras-2016}.

Combined, these surveys provide valuable datasets and allow us to characterize well the 
star formation potential of hundreds of cluster-forming clumps located throughout the Galaxy. 

\subsection{Identifying cluster-forming clumps}

The ATLASGAL dust continuum survey identified more than 10,000 clumps in the Galactic plane from their emission at 870 \mum ~\citep{Contreras-2013a, Urquhart-2014}. These clumps cover a large range of evolutionary stages from cold pre-stellar clumps to HII regions. To determine the physical and chemical properties of these star-forming clumps, MALT90 performed targeted observation of 16 spectral lines toward \nmalt~of the ATLASGAL clumps \citep{jackson-2013}, located between $-60$ to $+20$ degrees in longitude. The molecular lines observed by the MALT90 survey cover a wide range of excitation energies and critical densities providing a wealth of information about the chemical, physical and kinematic properties of these star-forming clumps. 

Because dust continuum surveys detect all the emission along the line of sight, it is possible that dust continuum peaks may be  due to physically distinct clumps superposed in projection but located at different distances; the kinematic information provided by MALT90 is key to determining whether this is the case or not. The MALT90 catalog \citep{Rathborne-2016} provides information regarding multiple dense gas molecular clouds along the same line of sight toward the dust continuum emission, differentiating those with single velocity components (S), or two velocity components (A and B). In this work we only used those clumps classified as single (S) that are clearly associated with a single velocity component from the dense gas emission along its line of sight. 

MALT90 data was also used to determine the kinematic distances toward the clumps (Whitaker et al. submitted). The kinematic distances were determined based on the flux-weighted average velocity of the emission ($v_{LSR}$) seen in the HCN(1-0), HNC(1-0), N$_2$H$^+$(1-0) and HCO$^+$(1-0) transitions. This average velocity was denominated as \textit{consensus velocity} of the clump \citep[see][]{Rathborne-2016}. Near and far kinematic distance disambiguation was determined using the H I absorption/emission \citep[e.g.,][]{kolpak-2003,Green-2011,Wienen-2015}. No distances were obtained for clumps within the galactic longitude range 350$^\circ < l < 15^\circ$. In this region, geometric effects and the large velocity dispersion observed in molecular clouds near the Galactic Centre lead to greater uncertainty in the distance determination. Also, for each clump a reliability factor for its kinematic distance was estimated. This factor has values ranging from -1 (poor) to 10 (most reliable) (see Whitaker et al. submitted). In this work, we only included clumps for which the kinematic distance reliability factor is $\geq4$, to avoid including clumps with kinematic distances not so well determined. 

Using the kinematic information provided by MALT90 we determined if two adjacent (in galactic longitude - $l$ - and latitude - $b$) dust continuum peaks actually arise from the same (more extended) physically coherent structure. We assumed that any column density peak located within $<1'$ (equivalent to 0.9 pc at 3 kpc) from each other, with a difference in their \textit{consensus velocity} $<2$ km s$^{-1}$ are part of the same physical clump. The values for distance and velocity separation were chosen to make sure that we are not selecting very extended low density emission. Using these parameters, we have identified \ncoherent~groups that are connected in the $(l,b,v)$ space.

For these groups of clumps, their areas and mean column densities are derived from the sum and average, respectively, of the clumps that comprise the groups. Table 1 summarizes the physical properties derived for the clumps. In this table the groups of clumps are listed using the MALT90 name of the clump with the highest column density. The column ``Notes" in this table shows the individual column density peaks that compose the group of clumps, indicated by their MALT90 name.

\subsection{Dust temperatures}

The dust temperatures of the clumps were derived from a simultaneous fit of the dust temperature and column density to the spectral energy distribution (SED) of the dust continuum emission detected by \textit{Hershel} at 160, 250, 350 and 500 \mum~and by ATLASGAL at 870 \mum~\citep{Guzman-2015}. This fit is done independently for each line of sight, taking into account differences in temperature across the clump better than previous studies \citep[e.g.][]{faundez-2004,Hill-2005}, in which a single temperature is assumed for the entire clump. However, for clumps with active star formation, it is possible that the column density is underestimated due temperature gradients along the line of sight. This bias might lead to an underestimation of the column density toward lines of sight associated with large temperature gradients.

In most cases, the consistency between the \textit{Herschel} and ATLASGAL flux measurements allowed for a good fit to dust temperature and column density. However, in some cases ($<10\%$ of the sources), in order to perform a good fit, it was necessary to remove some of the ATLASGAL data. In these clumps, the area affected (typically associated with low column density lines of sight) amounts to $\sim20\%$ of the total source area. The errors of the dust temperature and column density best fit were estimated from the projection of confidence ellipses on the dust temperature - column density plane of solutions. Calibration uncertainties in the \textit{Herschel} fluxes are expected to be of the order of 10\%, according to the Herschel SPIRE and PACS manuals and recommendations. The effect of larger (or smaller) calibration uncertainties are discussed in \citet{Guzman-2015}. For consistency,  we decided to adopt 10\% calibration uncertainties in this work.

As pre-stellar clumps evolve into stellar clusters, the energy radiation of the embedded proto-stars will start to heat its surrounding gas and dust.  In this scenario, we expect that cold ($<15$ K) dense ($>10^4$ cm$^{-3}$) clumps with no evident infrared emission represent the earliest stages of cluster formation with no active star formation yet. As the clumps evolve, the  UV radiation from the newly formed stars will start to  heat and eventually disrupt its environment, increasing the overall temperature of the gas and dust of the clumps. In the infrared, these clumps are identified in \textit{Spitzer} GLIMPSE and MIPSGAL images by their enhanced emission at 4.5 \mum, or with 8 \mum~or 24 \mum~point like or extended emission. 

Thus, because we expect an increase in the gas and dust temperature of a clump as it evolves, its dust temperature can be used as a proxy for its evolutionary stage.  Indeed, \citet{Guzman-2015} showed that the average dust temperature of the MALT90 clumps increased monotonically, from those that showed no IR emission, to those that showed obvious IR emission in \textit{Spitzer} GLIMPSE/MIPSGAL images. This suggests that the classification of the sources into `Quiescent', `Proto-stellar', `HII Regions' and `PDR' based on their IR emission done by MALT90 followed well the overall evolution of the clumps \citep[see][for details of this classification]{jackson-2013,Guzman-2015,Rathborne-2016}. Clumps with no infrared evidence of current star formation have dust temperatures of \mbox{$\sim$17$\pm$4 K}, while those with infrared signatures of active star formation typically have higher dust temperatures of \mbox{$\sim$20$\pm$4 K}. Clumps in later evolutionary stages, like those associated with HII regions, typically have dust temperatures of \mbox{$>26\pm5$ K}. Here, the uncertainties represent their internal variance in each group, and not uncertainty of measurements.
 
Since these results show evidence of a link between the clump dust temperature and the level of star formation activity, in this work we will use the dust temperature of the clump as a systematic measurement of the clump's evolution. The clumps in our sample have a wide range of dust temperatures, with a median dust temperature of 22.0 K, a minimum of 10 K and a maximum of 44 K (see Table 2).

\subsection{Clump masses and volume densities}

To derive the masses and volume densities of the clumps we used column densities and radii of the clumps derived from the \textit{Hershel} (160, 250, 350 and 500 \mum)~and  ATLASGAL (870 \mum) dust continuum maps and the kinematics distances from Whitaker et al (submitted). During the fit to the SED, the column density and dust temperature were simultaneously computed, ensuring that the measurements of the column densities are reliable by taking into account the differences of dust temperature within the clump \citep{Guzman-2015}.

The clump effective angular radii given by \citet{Guzman-2015} were obtained after convolving the emission at the higher angular resolutions (18" at 870 \mum) by a gaussian with a size of the beam of lowest angular resolution data (35" at 500 $\mu$m). To recover the most accurate measurement of the size of the clump, and to not underestimate the volume density, we computed the beam-corrected effective angular radius of the clump $\theta_{eff, deconv.}$. This radius was determined by de-convolving the effective radius of the clump as $\theta_{eff, deconv.} = \sqrt{\theta_{eff}^2-35"^2}$, where 35$"$ is the beam size. This size is also comparable to the sizes of the clumps previously determined by \citet{Contreras-2013a} and \citet{Urquhart-2014} using the ATLASGAL maps. The angular radius was converted then into physical radius using the distance to the sources via $R_{eff}=\theta_{eff, deconv.}D/206265$, with D the distance to the clump in pc and $\theta_{eff,deconv.}$ in arcsec.

The total mass for each clump was derived using the expression:
\begin{equation}
M=4.8\times 10^3 \left(\frac{N(H_2)_{mean}}{\textrm{gr~cm}^{-2}}\right) \times \left(\frac{A}{\textrm{pc}^2}\right) \times \mu_{H_2} M_\odot,
\end{equation}
\noindent where $N(H_2)_{mean}$ is the mean column density within the clump derived from the dust continuum emission assuming a gas-to-dust mass ratio of 100, A is its area, with $A=\pi R_{eff}^2$ and $\mu_{H_2} = 2.8$ is the mean molecular weight. The derived clump masses range from a few M$_\odot$ to $\sim10^5$ M$_\odot$,  with a median value of $1.0\times10^3$ \Msun (see Table \ref{tabla-resumen}).

The volume density, $n(\rm{H_2})$, was computed from the total mass and the volume ($V=\frac{4}{3}\pi R_{eff}^3$), assuming a spherical shape and a molecular weight of $\mu_{H_2} = 2.8$ (see Table \ref{tabla-resumen}). The volume density of clumps ranges from  $4.9\times10^2$ to $1.1\times10^7$ cm$^{-3}$. The median volume density is $\sim$ 6.8$\times 10^3$ cm$^{-3}$. 

\subsubsection{Virial mass and gravitational state of the clumps}

The gravitational state of each clump was determined by measuring its virial parameter, $\alpha$, defined 
as the ratio of the clump's virial mass to its total mass. The parameter $\alpha$ is commonly used to indicated whether or not a clump 
is gravitationally bound. If $\alpha >1$, then the clump contains more kinetic energy than gravitational energy, and is 
likely to expand unless confined by an external mechanism. In this case, the clump is considered to be unbound. A value 
of $\alpha <1$ suggests that the kinetic energy is insufficient to support the clump against its gravitational energy, and 
the clump will likely undergo collapse. In this case, in the absence of magnetic fields, the clump is considered to be bound.

The virial mass for each clump was calculated via \citep{Bertoldi-1992}:

\begin{equation}
M_{vir}=\frac{5}{Ga_1a_2}R_{eff}\sigma^2, 
\end{equation}
\noindent where G is the gravitational constant; $a_1=(1-p/3)/(1-2p/5)$, for $p<2.5$, which is a correction term that depends on the 
density gradient with power-law index $p$ \citep{Dunham-2010}, which we assume to be $p=1.8$ \citep{Mueller-2002}; and $a_2$ is a correction term 
that depends on the clump shape. For simplicity we assume the clumps are spherical, thus $a_2$ is 1 for all clumps. 
$R_{eff}$ is the effective radius of the clumps and $\sigma$ is the clump's velocity dispersion. 

The velocity dispersion for each clump was determined from the \nhp~or \hnc~line-widths obtained from the MALT90 molecular line emission \citep{Rathborne-2016}, as $\sigma=\Delta v/2.35$, assuming that the lines have Gaussian profiles. We used these molecules as they are good density tracers for the clumps, and the line-widths are not so severely affected by the dynamics of the source (e.g. outflows, infall profiles). If the \nhp~emission was detected toward the clump and the \nhp~hyperfine components were fitted, the \nhp~line-widths were used to determine the velocity dispersion of the clump. However, if the fit to the \nhp~hyperfine components was poor, we used instead the \hnc~line-widths to determine the velocity dispersion of the clump. For 91 clumps, there was no reliable detection of either \nhp~nor \hnc, thus, we did not compute the virial mass for these clumps and these clumps are not considered in the rest of the analysis. Thus, the total number of clumps included in our sample is \nmaltused.

The derived values for each clump's velocity dispersion,  virial mass, and virial parameter are summarized in Table \ref{tabla-resumen}. The velocity dispersion ranges from 0.2 to 5 km s$^{-1}$. We find typical values for the virial mass ranging from 21 M$_\odot$ to 3.8$\times 10^4$ M$_\odot$, with a median value of 1.2$\times 10^3$ M$_\odot$. The virial parameter ranges from 0.04 to 240 with a median of 1.1.

\subsubsection{Bolometric luminosity}

The bolometric luminosity of a clump provides information about the star formation activity within the clump. If the clump is in an early evolutionary stage with no embedded proto-stars, then the clump will have a low bolometric luminosity. On the other hand, if star formation has already started within the clump, then we expect that the radiation of the newly formed stars will increase its overall bolometric luminosity. Thus, the bolometric luminosity can also be used to evaluate the level of active star formation within a clump. 

We calculate the bolometric luminosity ($L_{\rm bol}$) extrapolated from the dust emission detected toward the clumps. Here we assume that all of the radiation coming from the clump can be described by a gray body with the opacity law given by $\kappa_\nu=0.14~(250~\mu{\rm m}/\lambda_{\mu{\rm m}})^{\beta}~~\text{cm$^2$~g$^{-1}$}$, and any internal optical or UV radiation is reprocessed by dust and emerges in the FIR/submm. The extrapolation of the fit to the dust thermal emission is expressed in terms of the column density, solid angle and dust temperature of the clumps via:

\begin{align}
L_{\rm bol}&=40\,L_\odot \left(\frac{T}{\rm 10~K}\right)^{5.7} \left(\frac{D}{\rm kpc}\right)^{2} \left(\frac{\Omega_s}{\rm arcmin^2}\right) 
\end{align}
\begin{align*}
\times \left(\frac{N_{\rm gas}}{\rm gr~cm^{-2}}\right)\times C(T,N_{\rm gas})~~,
\end{align*}
\noindent where D is the kinematic distance; $\Omega_s$ is the solid angle; 
$N_{\rm gas} = N_{\rm H_2} \mu_{\rm gas}m_{\rm H}$ is the mean gas column density, with $\mu_{H_2}=2.8$ the mean molecular weight for the molecular gas and m$_H$ the hydrogen mass;  T is the dust temperature; and $C(T,N_{\rm H_2})$ is a correction factor that takes into account opacity and dust temperature effects given by:
\begin{equation}
C(T,N_{\rm gas})=\left(1+1.6\times10^{-3} N_{\rm gas} T^{1.7}\right)^{-1}~~.
\end{equation}

\noindent Appendix A provides a more detailed explanation on the derivation of L$_{\rm bol }$. 

The bolometric luminosity of the clumps ranges from 4.9 to $1.4\times10^6$ L$_\odot$, with a median value of 4.9$\times 10^3$ L$_\odot$ (see Table \ref{tabla-resumen}).

\section{Discussion}

\subsection{Evolution of the physical properties}

We use the dust temperature of the clumps to trace the evolution in their physical parameters (mass, volume density, bolometric luminosity, and virial state). To do this, we separated the clumps into dust temperature bins of 5 K, between 15 and 30 K. Clumps with dust  temperature $\leq20$ K represent clumps in their earliest evolutionary stages, clumps with dust temperatures in the range \mbox{20 K $<\rm{T}_{dust}\leq$ 25 K} are those where star formation has already started, and clumps with dust temperatures $>25$ K are those at the most evolved evolutionary stages, usually associated with HII regions and PDRs. Table 2 shows the range of values for the full sample and for each of the dust temperature bins, and the percentage of clumps within each dust temperature bin.

\begin{figure*}
   \centering
   \includegraphics[trim=0cm 0 1cm 2cm,clip,width=.35\textwidth]{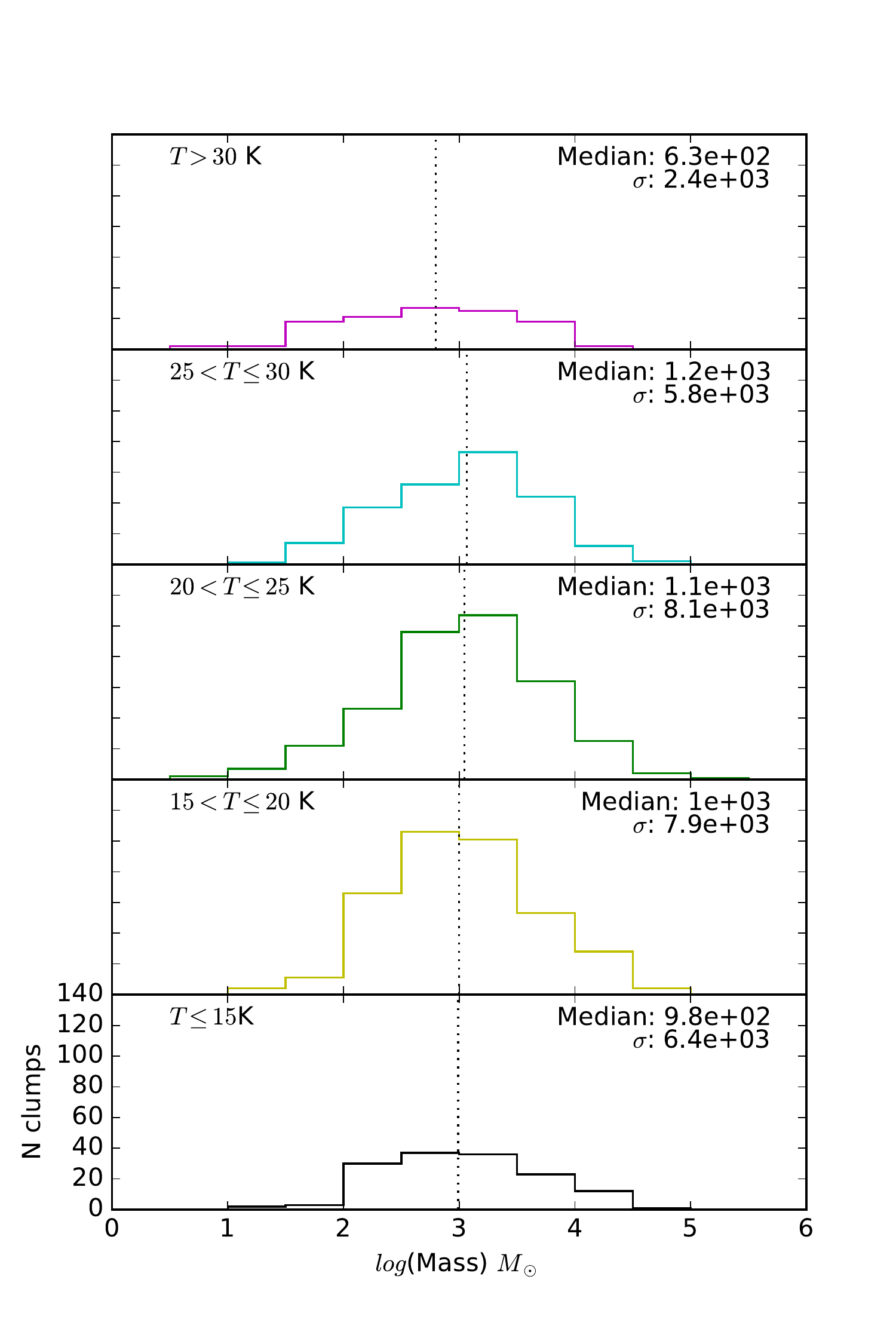}
   \includegraphics[trim=0cm 0 1cm 2cm,clip,width=.35\textwidth]{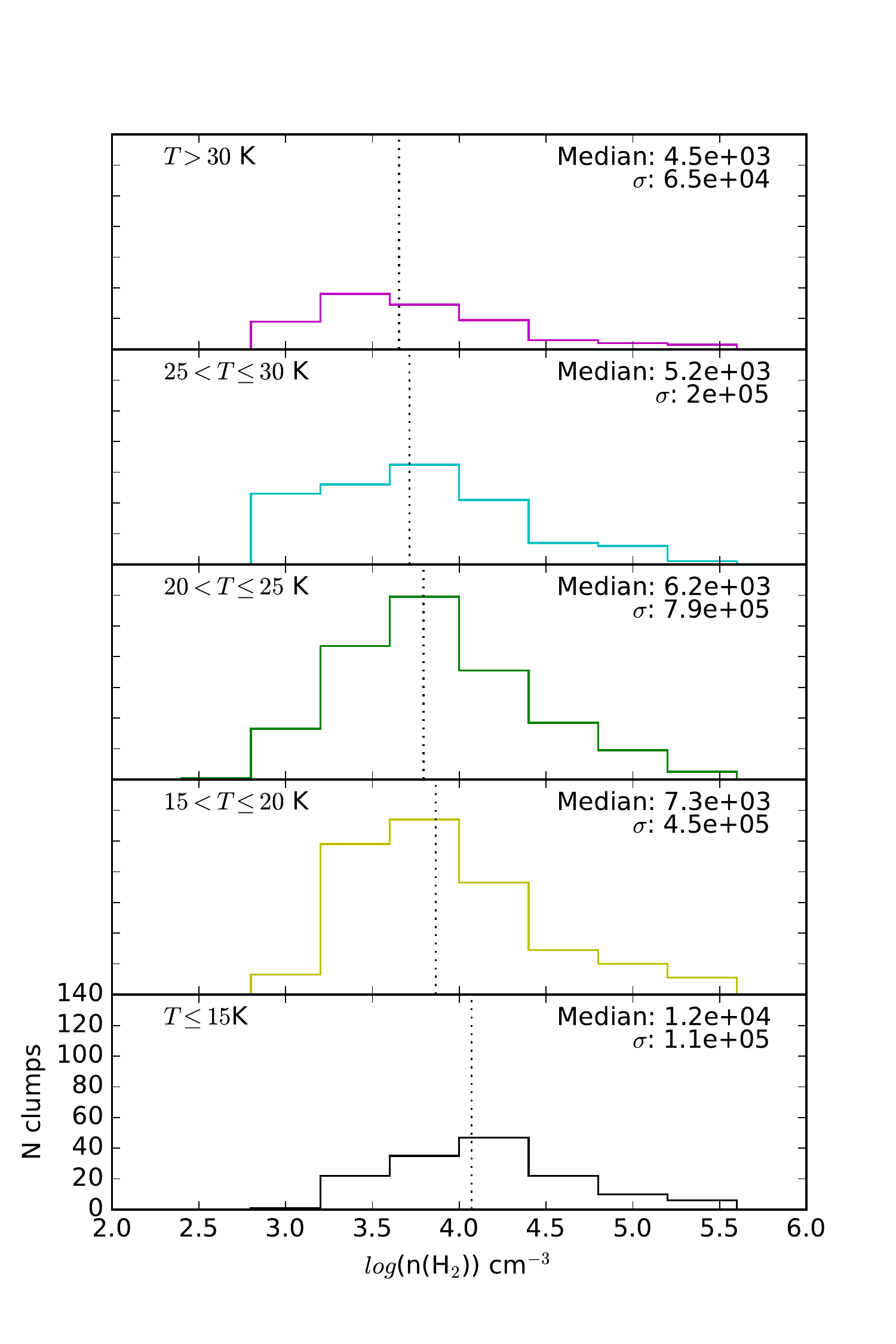}\\
   \includegraphics[trim=0cm 0 1cm 2cm,clip,width=.35\textwidth]{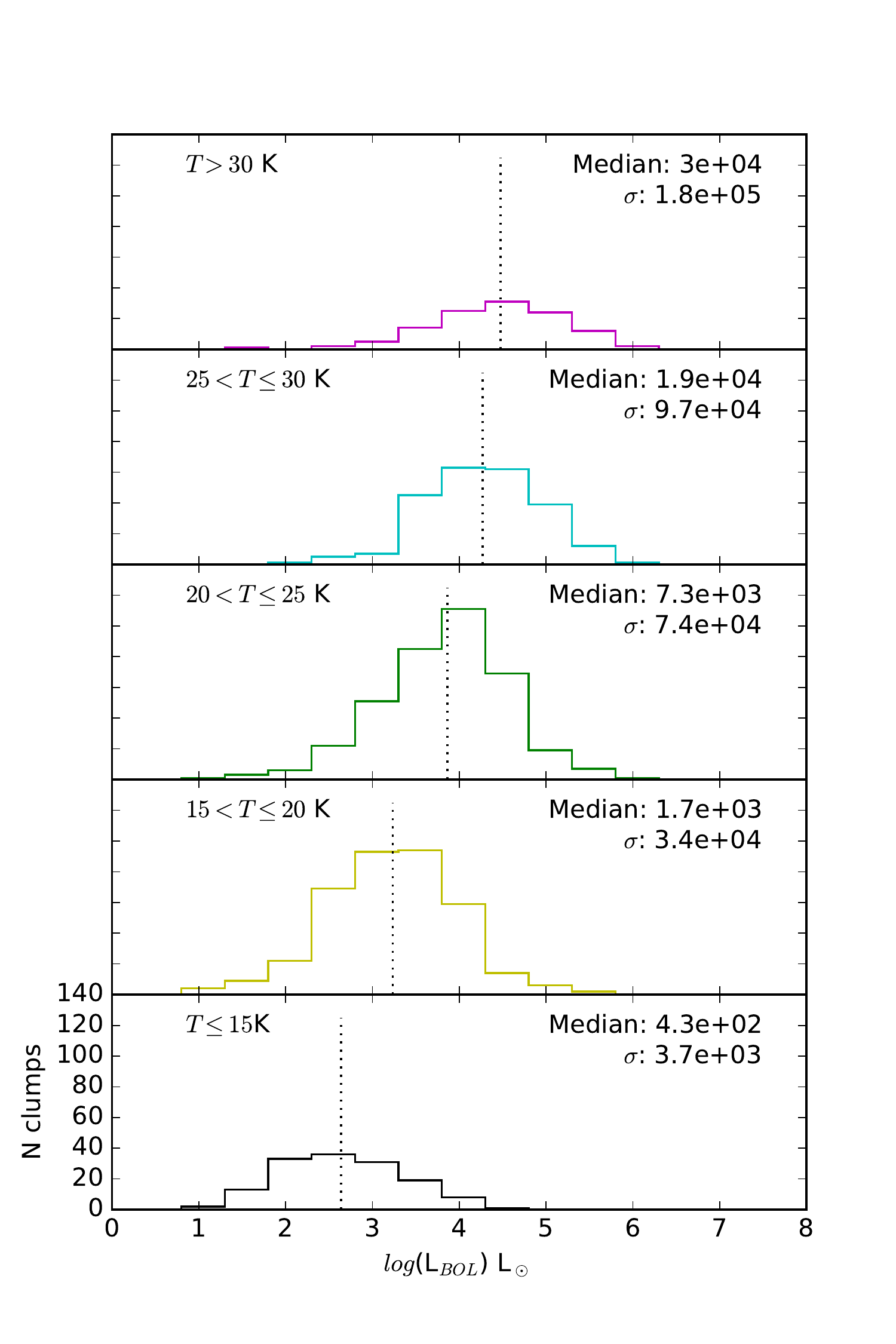}
   \includegraphics[trim=0cm 0 1cm 2cm,clip,width=.35\textwidth]{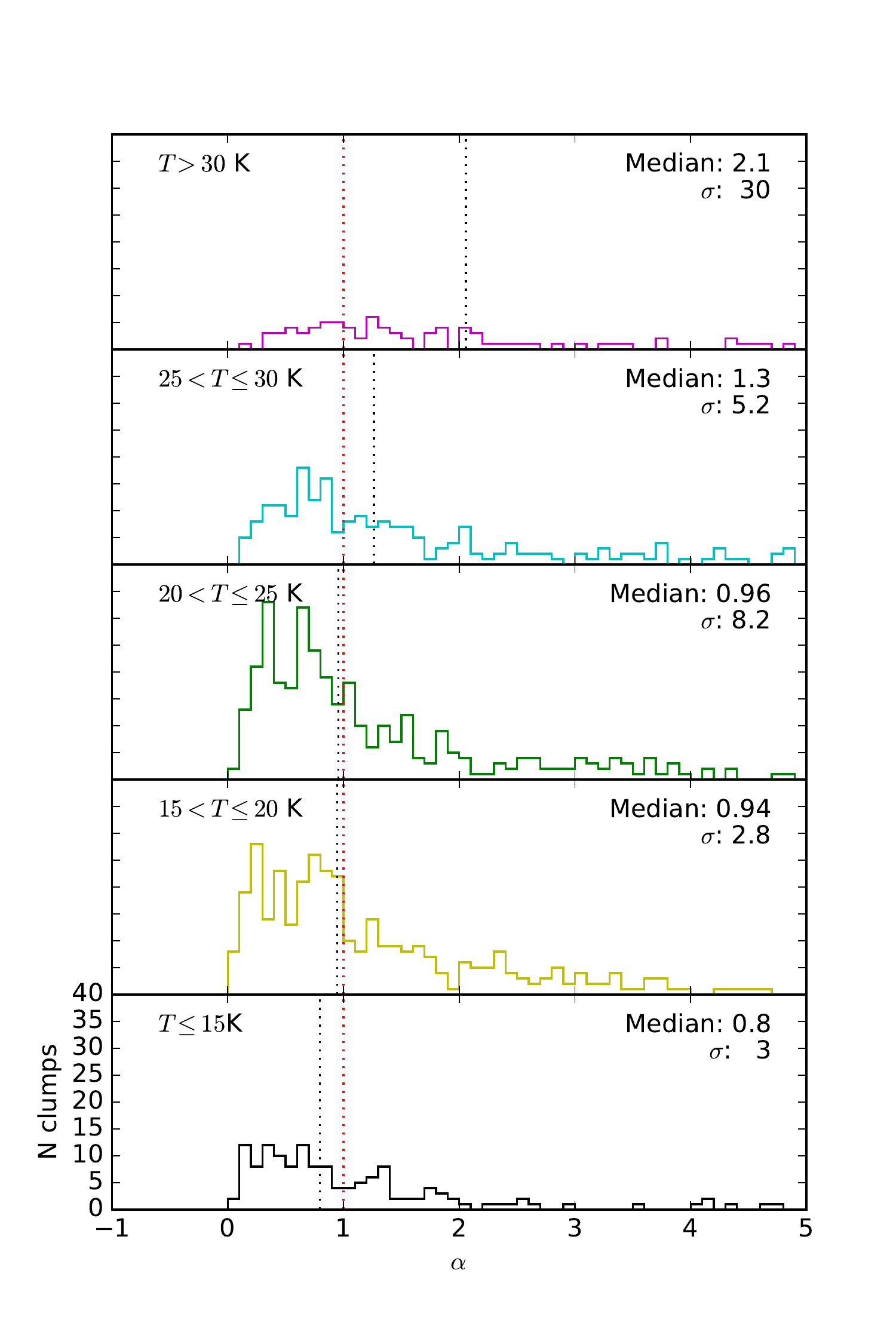}
      
   \caption{Histograms of clump mass, volume density, bolometric luminosity and virial parameter for the indicated dust temperature bins. In each plot the black dotted line shows the median value of the histogrammed quantity. This figure shows that the median value of the mass does not change significantly for clumps with different dust temperatures, having an median mass of $\sim10^3 \pm 10^2$ \Msun (upper left panel).  The upper right panel shows the volume density of the clumps for the same dust temperature bins. This figure shows that the volume density decreases for clumps with higher dust temperature, which is consistent with the evolution of the clumps in which clumps evolve thus diminishing their average volume density. The lower left panel shows the bolometric luminosity distributions for the different dust temperature bins. We see an increase of the bolometric luminosity with the dust temperature of the clumps, also consistent with the evolution of the clumps. The lower right panel shows the distribution of the virial parameter $\alpha$ for the different dust temperature bins.  In this plot the red dotted line shows $\alpha=1$. The virial parameter median value is $\la 1$ for clumps with dust temperature $<25$ K and increases to $\alpha>1$ for clumps with higher dust temperature. This is also consistent with the evolution of the clumps, where clumps are gravitationally bound at the onset of star formation and become unbound when they evolve.}
   \label{fig:mass-temp-bolo}
\end{figure*}

Figure \ref{fig:mass-temp-bolo} shows the mass, volume density, bolometric luminosity and virial state distribution of the clumps 
separated by their dust temperatures. The median mass for the clumps ranges from \mbox{$\sim 0.6 \times 10^3$ to $1.2 \times 10^3 $ \Msun}~for all the evolutionary stages, and is similar to the median value of the full sample of $1.0\times10^3$ M$_\odot$. 

For each dust temperature bin, we computed a Kolmogorov-Smirnov (KS) test of the mass distribution with respect to the full sample of clumps. Table 3 shows the results of the KS value and probability for each physical parameter. We find that for clumps with $\rm{T}_{dust}\leq 30~K$, the mass distribution in each bin is similar to the mass distribution of the full sample. For clumps with $\rm{T}_{dust}>30$ K the KS test shows that the mass distribution is different from the mass distribution of the full sample, which might be due to the lower number of clumps with these dust temperatures. The median values of the mass and the KS test suggest that the mass distribution of the clumps remains fairly constant regardless of their dust temperature, thus, we are tracing similar mass ranges for all evolutionary stages. 

KS tests on the mean volume density distribution shows that the density distribution for each dust temperature bin is different from the distribution of full sample (see Table 3). This suggests that the volume density of a clump changes with its evolution. We found that the median volume density of clumps decreases with the dust temperature. This is consistent with the scenario where at later evolutionary stages, and thus higher dust temperatures, the clumps have already formed protostars and have started to expand, thereby increasing their size and decreasing their mean volume density. 

The bolometric luminosity of the clumps also increases with the dust temperature of the clumps. This change is confirmed by the KS test between the distribution of bolometric luminosity for each dust temperature bin and the full sample (see Table 3). We found that the dust bolometric luminosity of the clumps increases with dust temperature as expected, given that the bolometric luminosity is a steeply increasing function of the dust temperature of the clump. This is also consistent with the fact that clumps with higher temperature likely already have embedded protostars, which are heating their environment, increasing the overall bolometric luminosity of the clumps. 

We now analyze the virial parameters of the clumps. KS tests show that the distributions for clumps with dust temperature $\leq15$ K and clumps with $\rm{T}_{dust}>25$ K are different from the distribution exhibited by the full sample. However, for clumps with dust temperature between 15 K $< \rm{T}_{dust} \leq$ 25 K the distribution of the virial parameter is similar to the distribution of the full sample. Thus, the virial parameter also increases with the evolution of the clumps. For clumps with dust temperatures $\leq25$ K, the median values of $\alpha$ are $ \la 1$, while for clumps with dust temperatures $>25$ K, the median value of $\alpha$ is greater than one. This trend is also consistent with the evolution of the clumps. When the clumps have a lower dust temperature, they are more bound. Once star formation has begun, the protostars will start to heat and ionize their environment. The temperature of the clump will increase, and the protostar will start dissipating its natal gas. The feedback from the newly formed stars (e.g. outflows and winds) will increase the turbulence and therefore the virial mass. At this stage the clump as a whole will become more gravitationally unbound.  

\subsection{Free-fall time and dynamical time of the clumps}

We determined the degree of gravitational instability of the clumps as a function of the clumps masses. For this we determined the free-fall time, $t_{ff}= \sqrt{3\pi/ 32 G\rho}$, and the dynamical crossing time or dynamical time, $t_{dyn}=R_{eff}/\sigma$ \citep{mckee-ostriker-2007,Tan-2006}, for the clumps. A clump will become gravitationally unstable if the free-fall time is less than its dynamical time. Moreover, since the ratio $t_{eff}/t_{dyn} \propto \sqrt{\alpha}$, as $t_{ff}/t_{dyn}$ goes below 1, clumps will become sub-virial.

The free-fall time is relatively insensitive to the value of the clump mass, but the dynamical time is sensitive to the value of the mass. Because gravitational collapse will start for those clumps for which the free-fall time is less than the dynamical time, this result suggests that clumps with higher masses are more susceptible to gravitational instabilities and they evolve faster than their lower mass counterparts (see Figure \ref{fig:times}). This is in agreement with the theoretical picture for high-mass star formation, where clumps with higher mass evolve faster, quickly disrupting their natal environment \citep[e.g.,][]{Tan-2014}.

While most of the clumps have $t_{ff}/t_{dyn}<1$ and thus are likely to collapse, some clumps have $t_{ff}$ greater than their $t_{dyn}$. These clumps can be divided into two categories: (1) Clumps where star formation have already begun (dust temperature $>$20 K), with high virial parameter ($<\alpha> = 5.6\pm1.8$) due to activity from the stars (e.g. outflows, winds that increase the turbulence within the clump) and thus high $t_{ff}/t_{dyn}$ ($<t_{ff}/t_{dyn}> = 1.4\pm0.2$ ), and (2) Clumps in early stages of evolution (dust temperature $\leq$20 K) that have low volume density ($<n(\rm{H_2})>=6.4\pm0.5\times10^3$ gr cm$^{-3}$), low bolometric luminosities ($<L_{bol}>=4.7\pm0.9$ L$_\odot$), and are unbound ($<\alpha>=5.0\pm1.3$); thus, they might not collapse to form stars.

To determine the potential of collapse for the clumps that are in an early evolutionary stage we analysed the ratio of the free-fall time to the dynamical time ($t_{ff}/t_{dyn}$) for the clumps with dust temperatures $\leq20$ K. We determined the fraction of clumps that have $t_{ff}/t_{dyn}<1$ compared to the total number of clumps for any given volume density. For volume densities $<2.5\times 10^4$ cm$^{-3}$, 82\% of the clumps in our sample have values of $t_{ff}/t_{dyn} < 1$ and thus they are prone to gravitational collapse, while for volume densities $>2.5\times 10^4$ cm$^{-3}$ the number of clumps that have values of $t_{ff}/t_{dyn} < 1$ increases to a 97\%. This value of volume density agrees with previous studies of solar neighbourhood clouds, which suggest that star formation only occurs when the volume density exceeds a threshold density of $\sim10^4$ cm$^{-3}$ \citep[e.g.,][]{Lada-2012}.

\begin{figure*}
   \centering
   \includegraphics[trim=0cm 0 0cm 0cm,clip,width=.7\textwidth]{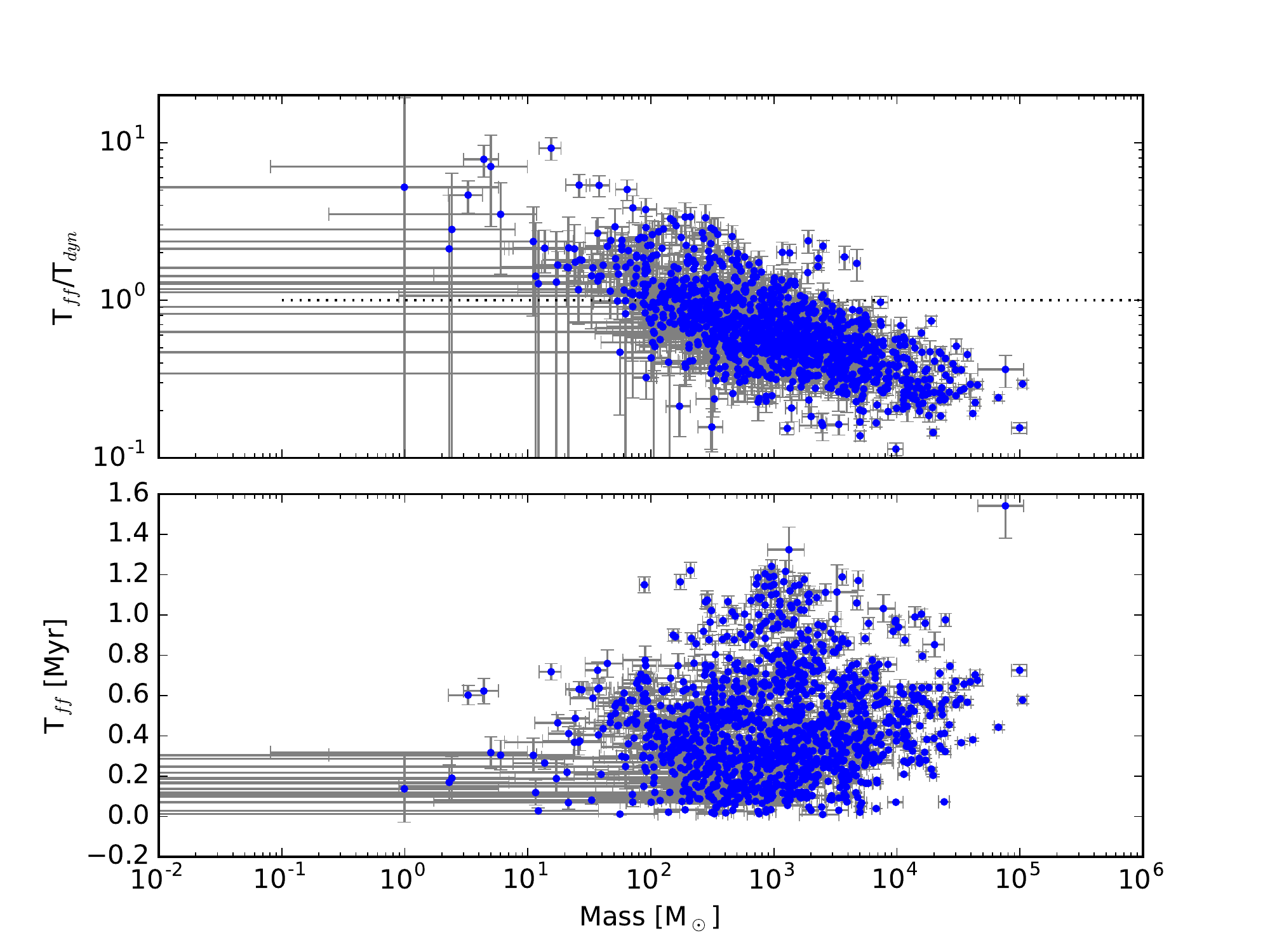}
   
\caption{Upper panel: Ratio of the free-fall to the dynamical time versus mass of the clumps, the dotted line marks where the free-fall times is equal to the dynamical time. The crosses show the uncertainties in the measurements, which are mostly due to uncertainties in the kinematic distance. In this plot we can see that the clumps with higher mass evolve faster. Lower panel: Free fall time versus mass of the clumps. This plot shows that while there is a large dispersion in the free fall time, it remains fairly constant with the mass of clump.}
   \label{fig:times}
\end{figure*}

\begin{figure*}
   \centering
    \includegraphics[trim=0cm 6cm 0cm 0cm,clip,width=.41\textwidth]{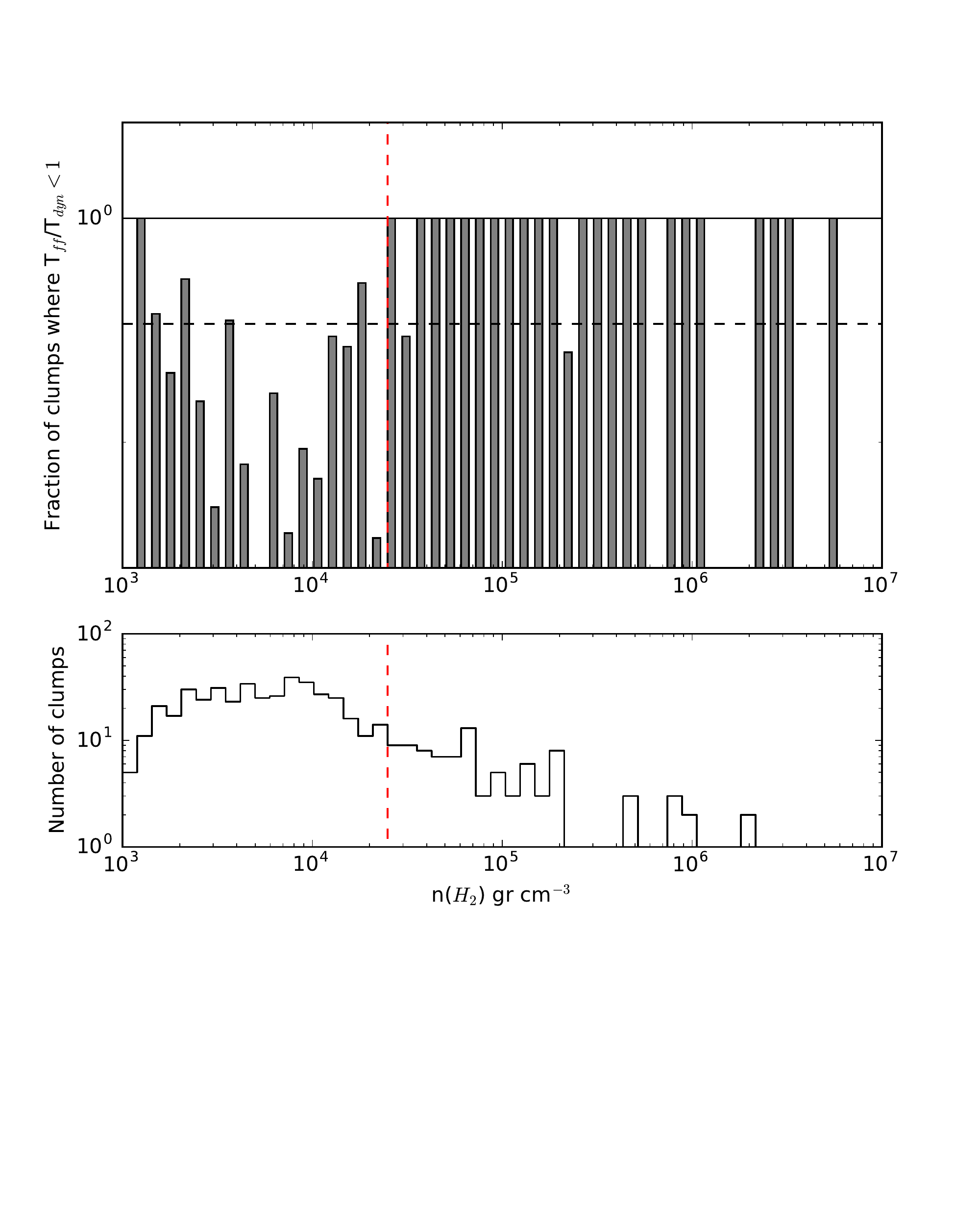}
       \includegraphics[trim=0cm 0 0cm 0cm,clip,width=.5\textwidth]{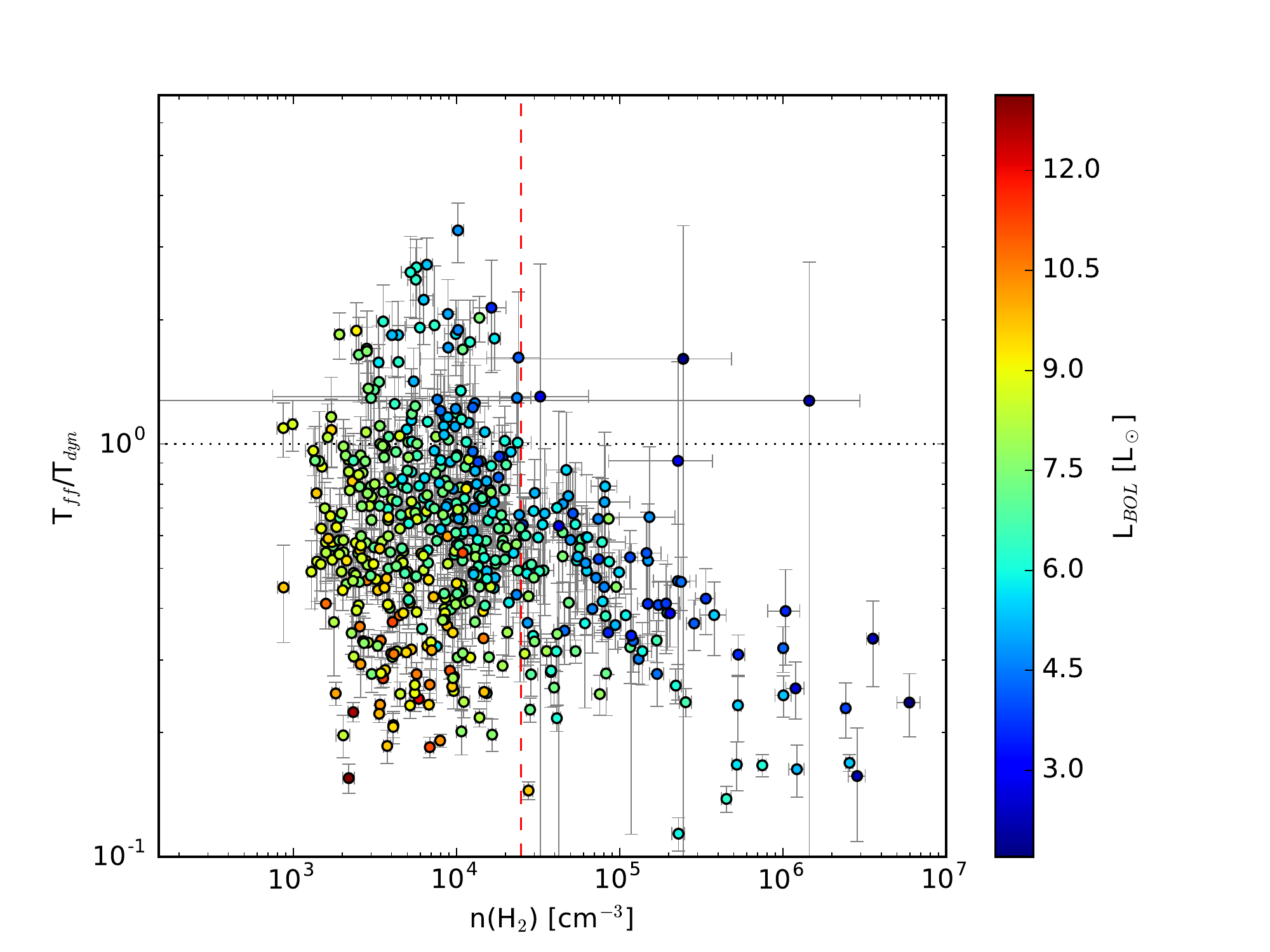}

\caption{Left, upper panel: Fraction of clumps with $t_{ff}/t_{dyn}<1$ as a fraction of the total number of clumps. The solid black line marks the ratio $=1$ and the dashed black line marks the ratio equals 0.9.  Left, lower panel: Number of clumps as a function of volume density. Right panel: Ratio of the free-fall time over dynamical time ($t_{ff}/t_{dyn}$) versus volume density for the clumps with dust temperature $\leq20$ K. Colors shows the bolometric luminosity of the clumps. Gray crosses show the uncertainty in the measurements. The dotted black line marks $t_{ff}=t_{dyn}$, clumps below this line are more susceptible to instability and collapse. In all the plots the dashed red lines mark the volume density for which the ratio starts to be $>90$\%, $n(H_2)=2.5\times10^4$ cm$^{-3}$, above which nearly all the clumps are prone to collapse.}
   \label{fig:times}
\end{figure*}

\subsection{High-mass star and cluster-forming potential of the clumps}

\begin{figure*}
   \centering
   \includegraphics[width=0.9\textwidth]{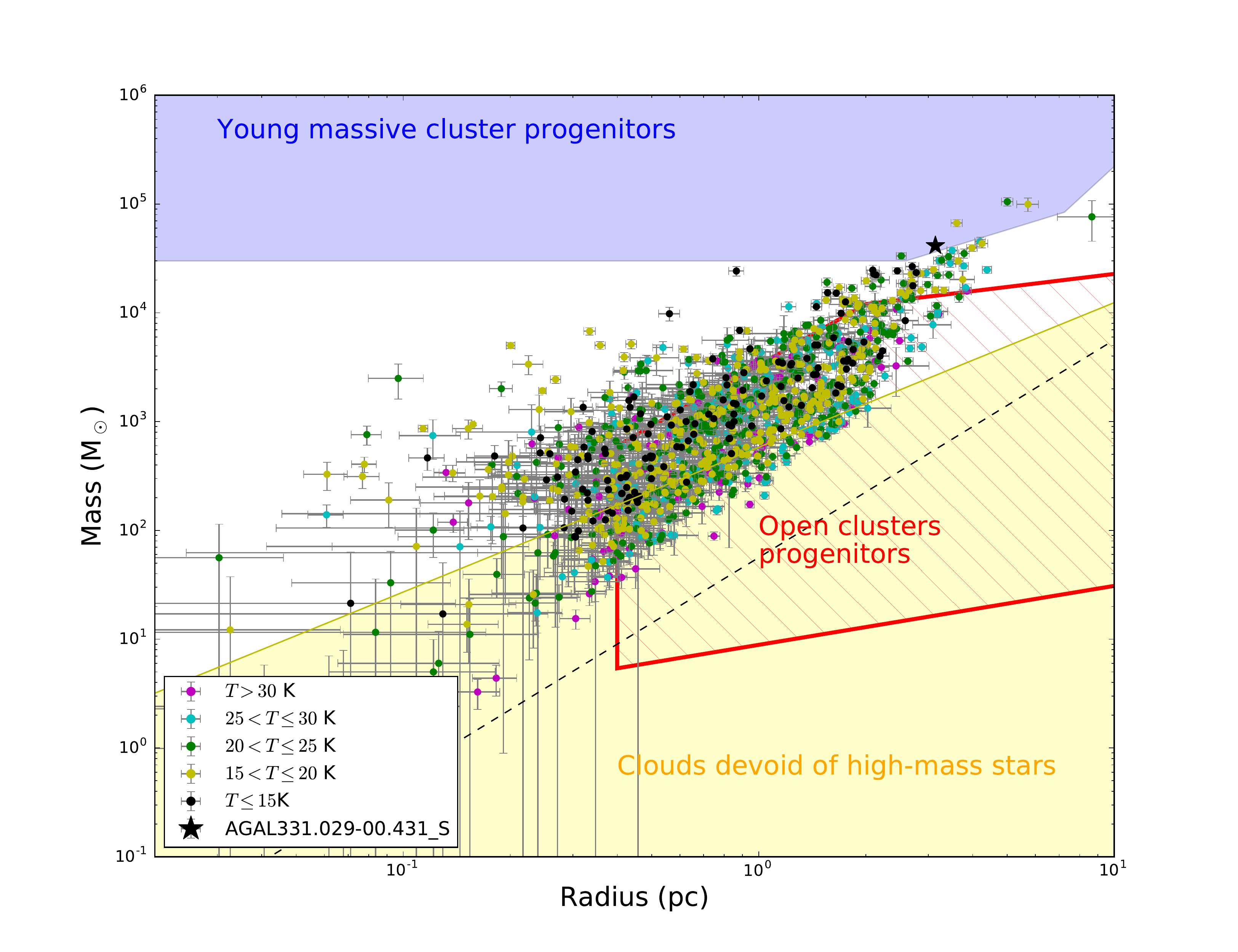} 
   \caption{Mass-size relationship between the clumps. In this plot, the clumps are color coded by dust temperature bins. Uncertainties in the measurements are shown with gray crosses. These uncertainties are dominated by the errors in the kinematic distance determination. The red dotted line shows the fit to the sample, and the black dotted lines shows the ATLASGAL nominal sensitivity of 0.25 mJy/Beam, which corresponds to 2$\times 10^{21}$ cm$^{-2}$. The yellow region highlights the parameter space for clouds that empirically have been shown that do not harbor high-mass stars \citep{kauffmann-2010}. Most of the clumps in our sample are above this limit, suggesting they will evolve into clusters harboring high-mass stars. The red hatched region shows the parameter space for clump progenitors of open clusters \citep{Portegies-2010}, assuming a star-forming efficiency of 30\%. 69\% of the clumps lie within this region. The blue region highlights the parameter space for YMC progenitors \citep{Bressert-2012}. There are just a few clumps meeting these criteria, which suggests that YMCs are rare in the plane of the Galaxy. Among the YMC progenitors one clump (marked with a black star in the plot) appears to be at the earliest stage of evolution.}
   \label{fig:mass-size}
\end{figure*}

To determine which clumps have sufficient mass and density to form stars and stellar clusters, we analyzed the mass and radius of the clumps. Figure \ref{fig:mass-size} shows the mass-radius relationship for the clumps, colour coded by dust temperature bins. By fitting the mass and radius of the clumps, we found that they follow the relationship M $= 1.20 \pm 0.01 \times10^3 ~R_{eff}^{2.2 \pm 0.01}$, suggesting that the clumps have nearly constant column density.  This result is slightly steeper than the mass size relationship of $\rm{M}_{dust}$ $\propto R_{eff}^{1.67}$ found by \citet{Urquhart-2014}, based on masses derived for a sample of ATLASGAL clumps assuming a constant dust temperature of 20 K.

We compared the clump masses and sizes with the empirical threshold found by \citet{Kauffmann-2010b} for clouds in which high-mass stars are usually found. While the clumps below this empirical threshold might or not form stars, we do expect that all the clumps above this threshold will eventually form at least one high-mass star. In this work we used a factor of 580 instead of the original 870 value. This is to account for the value of the opacity used in the column density calculation. The majority (78\%) of the clumps in our sample lie above this limit, suggesting that they have sufficient mass and density to harbor or eventually form at least one high-mass star. 

We also analyzed the potential of clumps to evolve into open clusters. For this we assumed that the clumps have a 30\% star-formation efficiency. Figure \ref{fig:mass-size} shows the parameter space in a plot mass versus radius allowed for open cluster progenitors \citep{Portegies-2010}. We found that $\sim$67\% of the clumps in our sample have masses and sizes similar to those expected for the progenitors of open clusters. 

The parameter space for clumps that have the mass and density of young massive cluster (YMC) progenitors is located in the upper end of this mass-size relationship. Five clumps within our sample lie in this region. Four of these 5 clumps have dust temperatures $\geq20$ K, and present some infrared emission in the \textit{Spitzer} maps, suggesting that star formation may have already begun within them. The small number of clumps that fulfil the requirements of YMC progenitors reinforces the idea that massive proto-clusters are rare in the Galaxy.

In Table 2 we provide a summary of the physical properties of the clumps that have the characteristics of progenitors of clusters harbouring high-mass stars (HMPC), and clusters that may evolve into forming open clusters (OC) and YMCs.

\subsection{Cold starless clumps}

We found that 19\% of the clumps have dust temperatures $\leq20$ K and densities $\geq10^4$ cm$^{-3}$ and thus have the properties of cluster-forming clumps in early evolutionary stages. These clumps are gravitationally bound: the median value of their virial parameter is $\alpha$ $\sim$0.8, suggesting they are likely to form stars. 
The mass-radius relationship for these cold clumps shows that the majority (88\%) have physical characteristics consistent with clump progenitors of high-mass stars.  A large fraction (62\%) of the clumps are candidates for open cluster progenitors, with the majority of these open cluster progenitors potentially harbouring high-mass stars. 

Within this sample of cold clumps, there are \ncoldclumps~clumps that lack emission in the 70 \mum~\textit{Herschel} maps \citep[denominated as far-IRDC by][for their lack of emission at far-IR wavelengths]{Guzman-2015}. These clumps have a median dust temperature of 14 K, a median mass of  $\sim10^3$ M$_\odot$, and a median volume density of $\sim1 \times 10^4 $ cm$^{-3}$. Consequently, they have the mass and density necessary to start forming stars.  All but one of these clumps have physical properties consistent with clumps that will eventually form high-mass stars. Thus, this suggests that these cold, dark at 70 \mum~clumps are in the very earliest stage of evolution, and thus they are excellent candidates to study the initial conditions of high-mass star formation.

We found only one cold clump, \ymca, with characteristics consistent with a YMC progenitor. Clump \ymca,  composed of the MALT90 sources \ymca~and \ymcb, has a mean dust temperature of 14 K. The H I data toward this clump suggest that it is located at a far kinematic distance of 10 kpc (Whitaker et al., submitted). \citet{Wienen-2015} located this clump at its near kinematic distance of $\sim$4 kpc. If we locate this clumps at this distance, then, it would have a mass of $6.6\pm0.8 \times 10^3$ M$_\odot$, a density of $3.9\pm0.3 \times 10^4$ cm$^{-3}$, and a dust bolometric luminosity of 4.9$\pm0.9 \times 10^3$ L$_\odot$. However, this clump does not appear as an infrared dark cloud in the \textit{Spitzer} GLIMPSE images, in agreement with what is expected for a clump located at a distance of 10 kpc. At this distance the clump has a mass of  4.1$\pm0.2 \times 10^4$ M$_\odot$, a density of 7.9$\pm0.3 \times10^3$ cm$^{-3}$, and a dust bolometric luminosity of 3.1$\pm0.6\times 10^4$ L$_\odot$. 

\begin{figure*}
\begin{center}

\includegraphics[trim=2cm 5cm 2cm 3.cm,clip,width=0.4\textwidth, angle=-90]{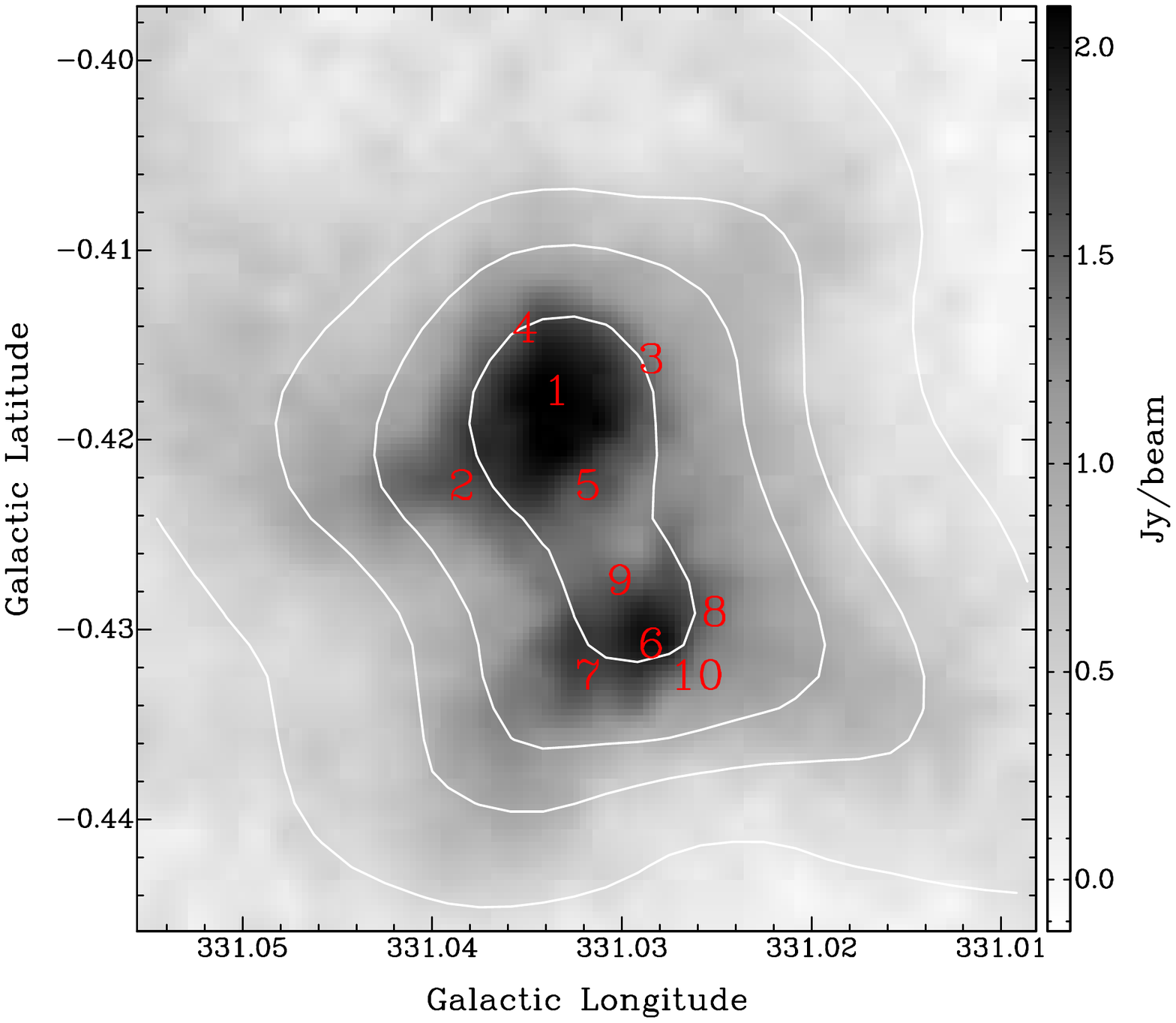}
\includegraphics[trim=2cm 5cm 2cm 3.cm,clip,width=0.4\textwidth, angle=-90]{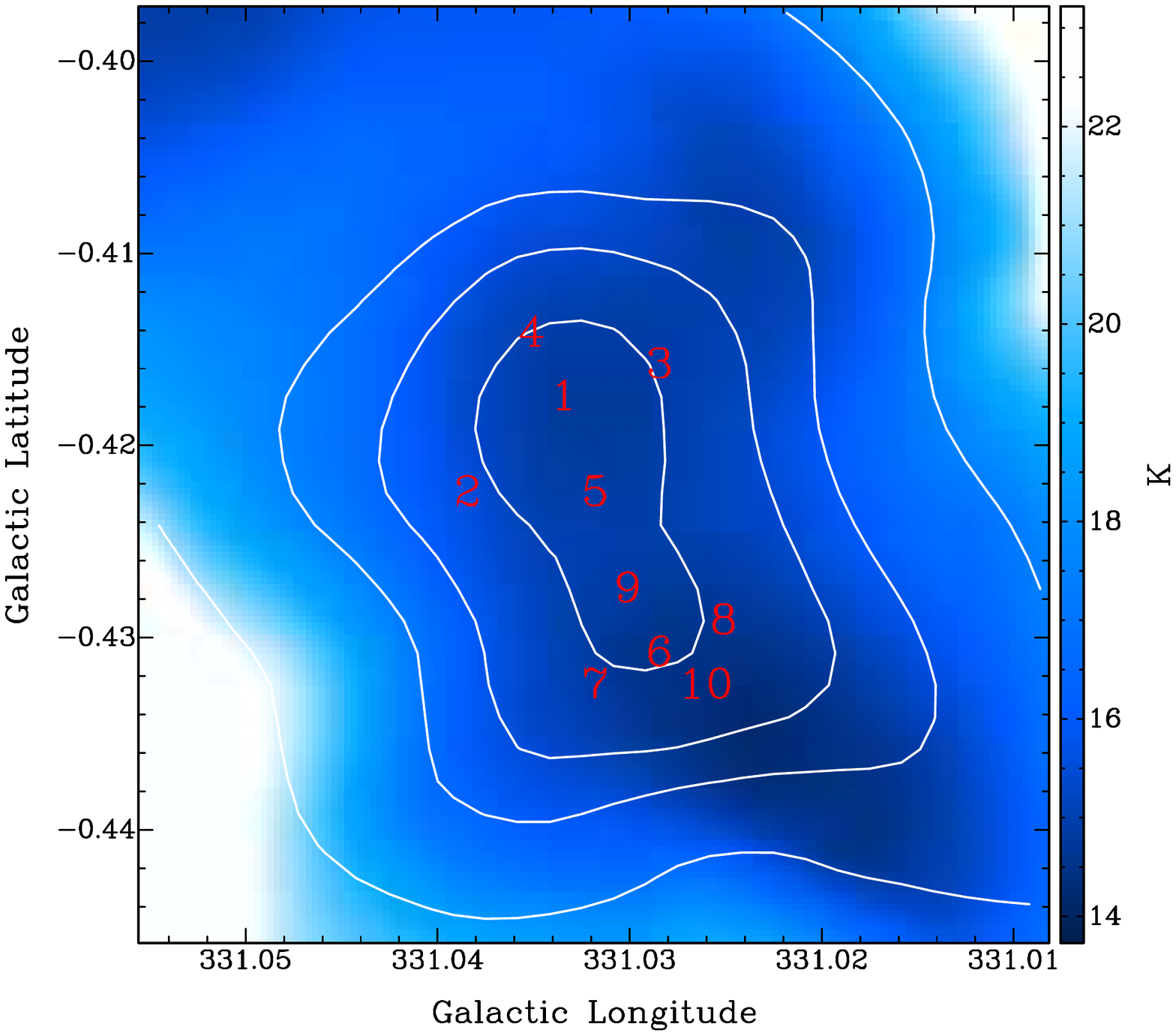}\\
\end{center}
\includegraphics[width=0.2\textwidth]{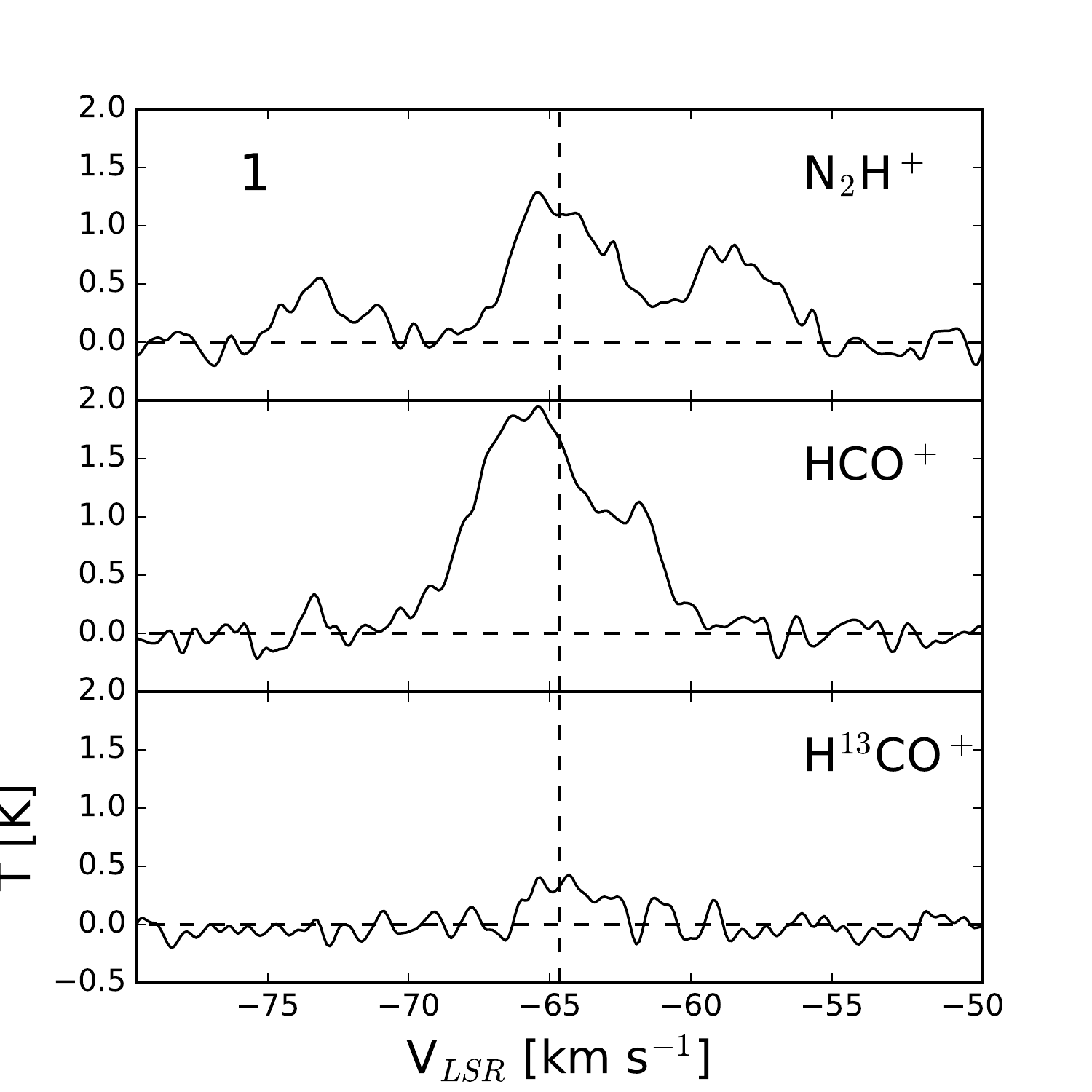}\hspace{-0.3cm}
\includegraphics[width=0.2\textwidth]{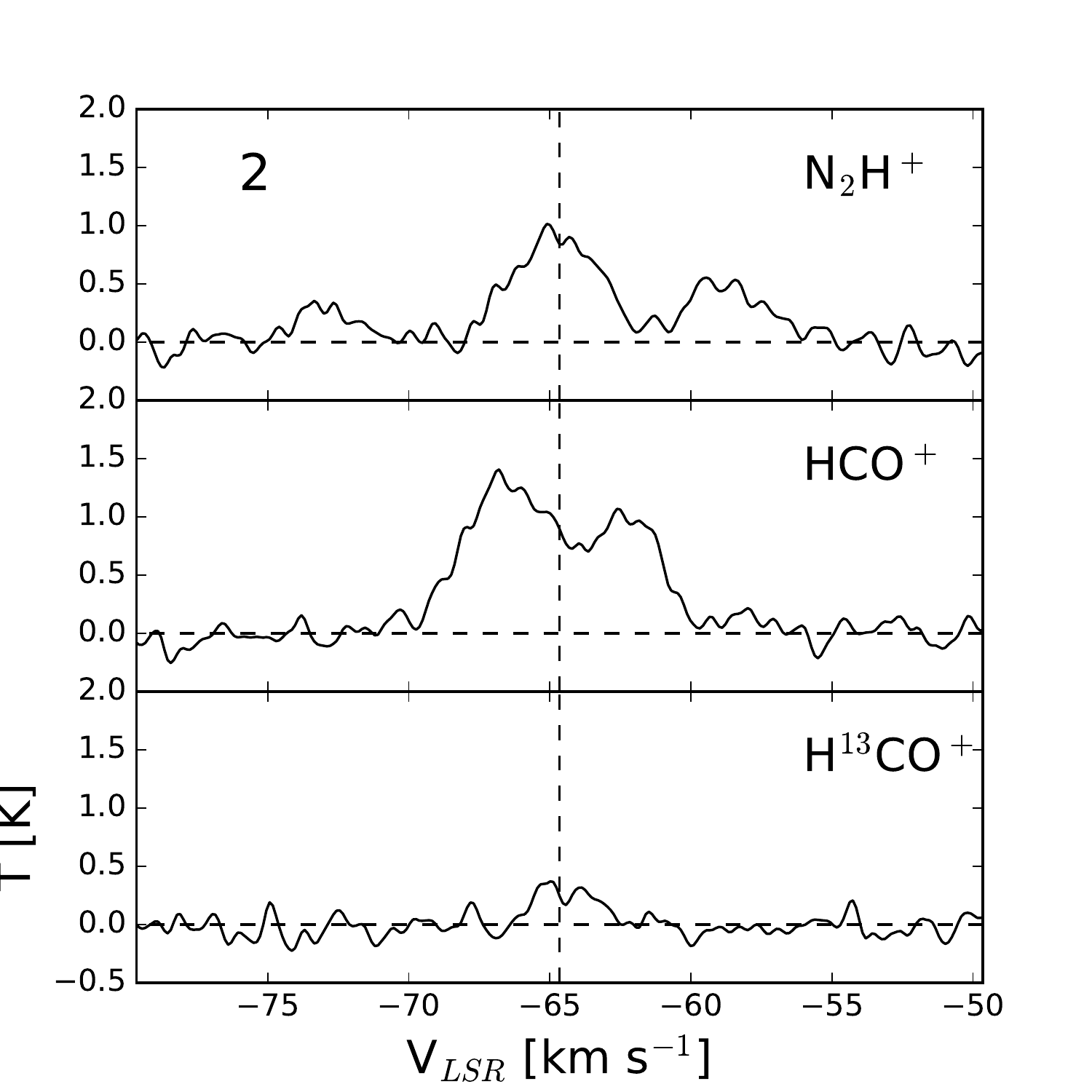}\hspace{-0.3cm}
\includegraphics[width=0.2\textwidth]{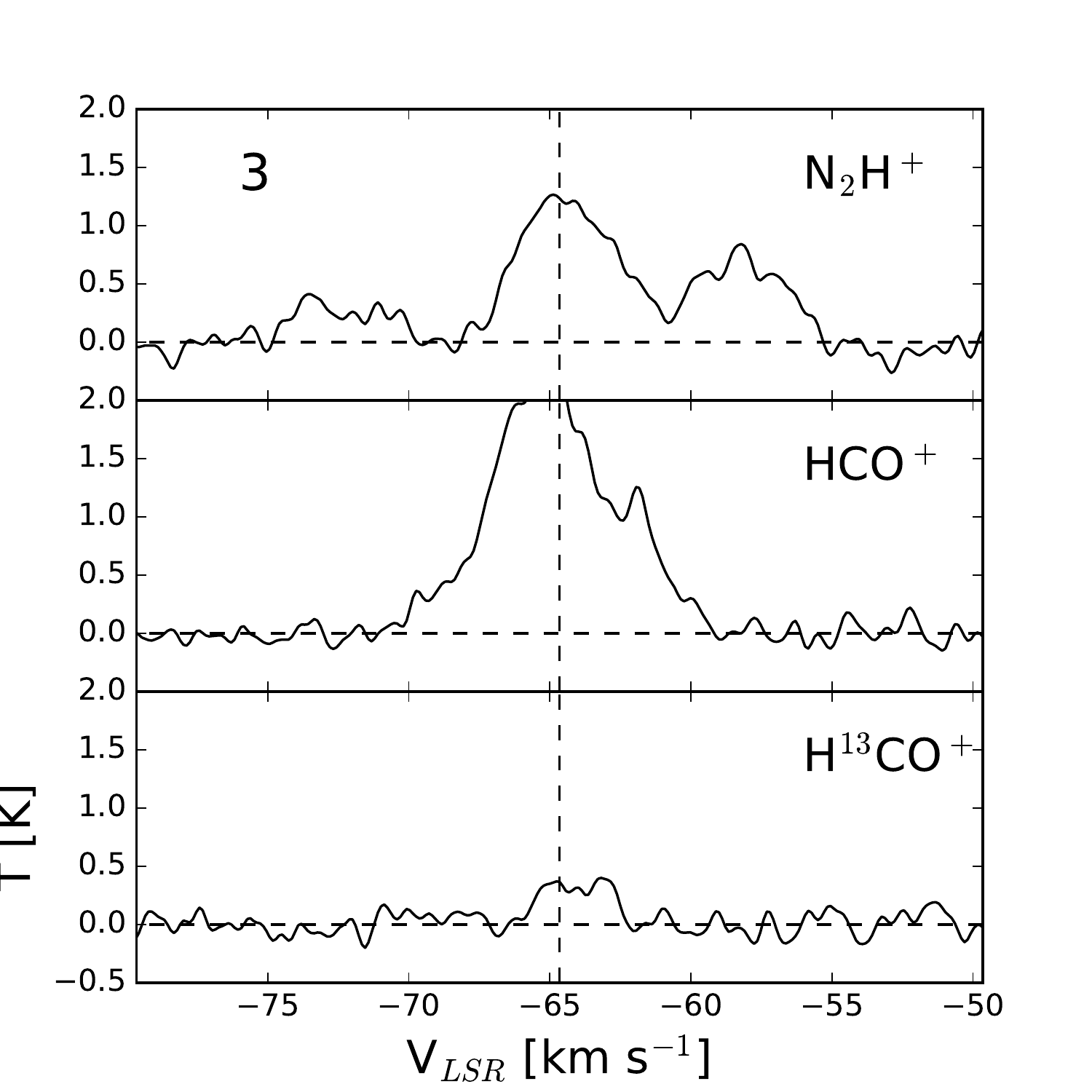}\hspace{-0.3cm}
\includegraphics[width=0.2\textwidth]{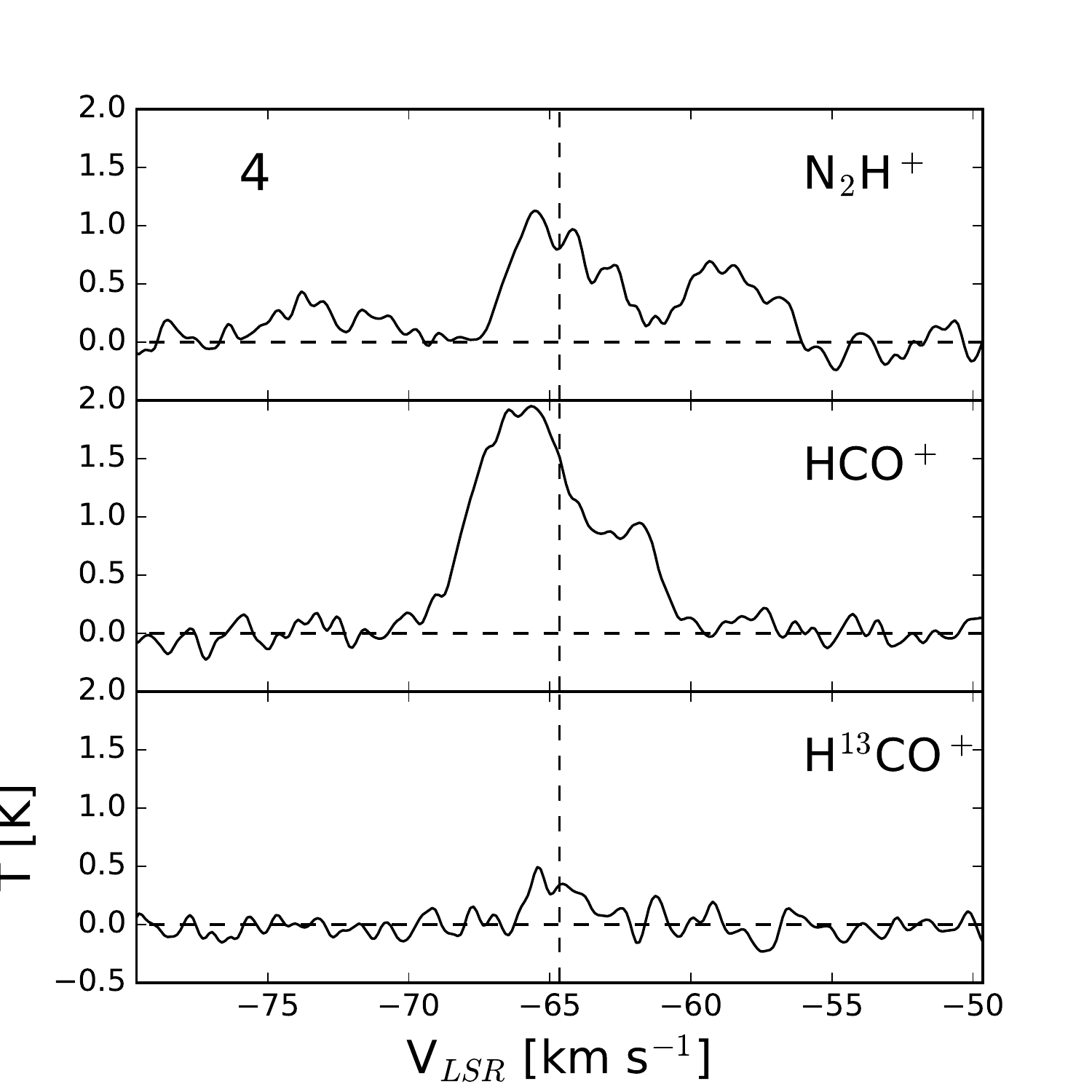}\hspace{-0.3cm}\\
\includegraphics[width=0.2\textwidth]{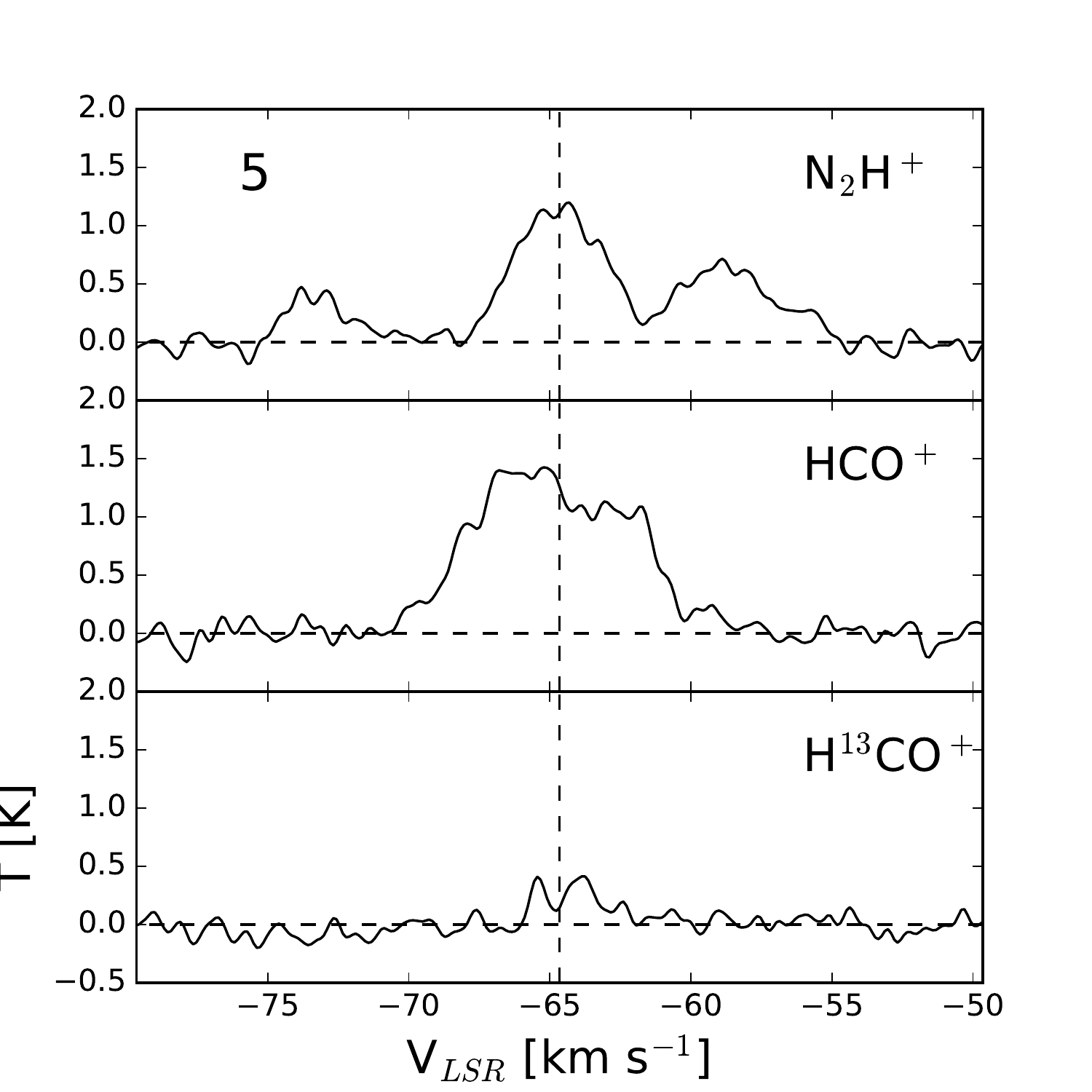}\hspace{-0.3cm}
\includegraphics[width=0.2\textwidth]{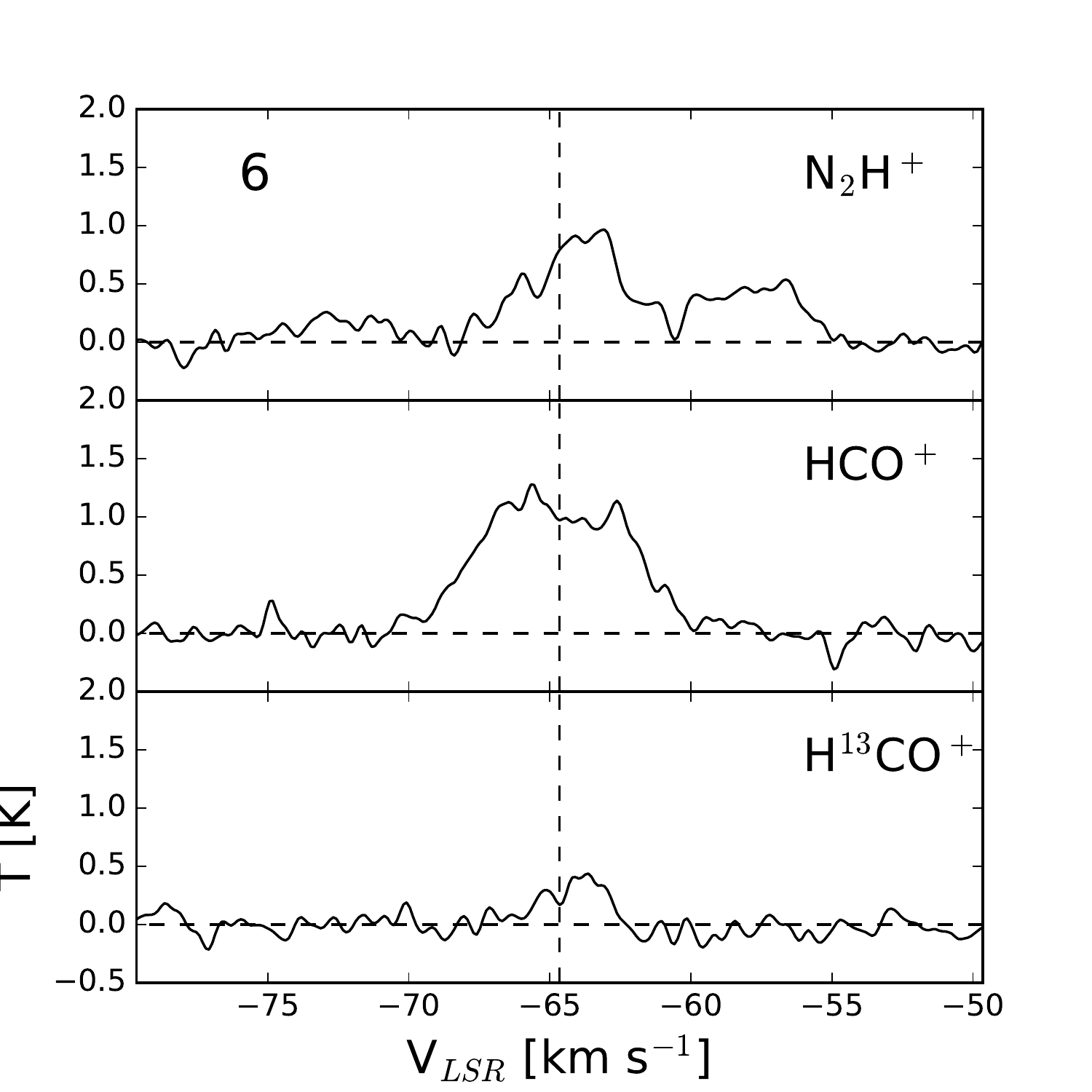}\hspace{-0.3cm}
\includegraphics[width=0.2\textwidth]{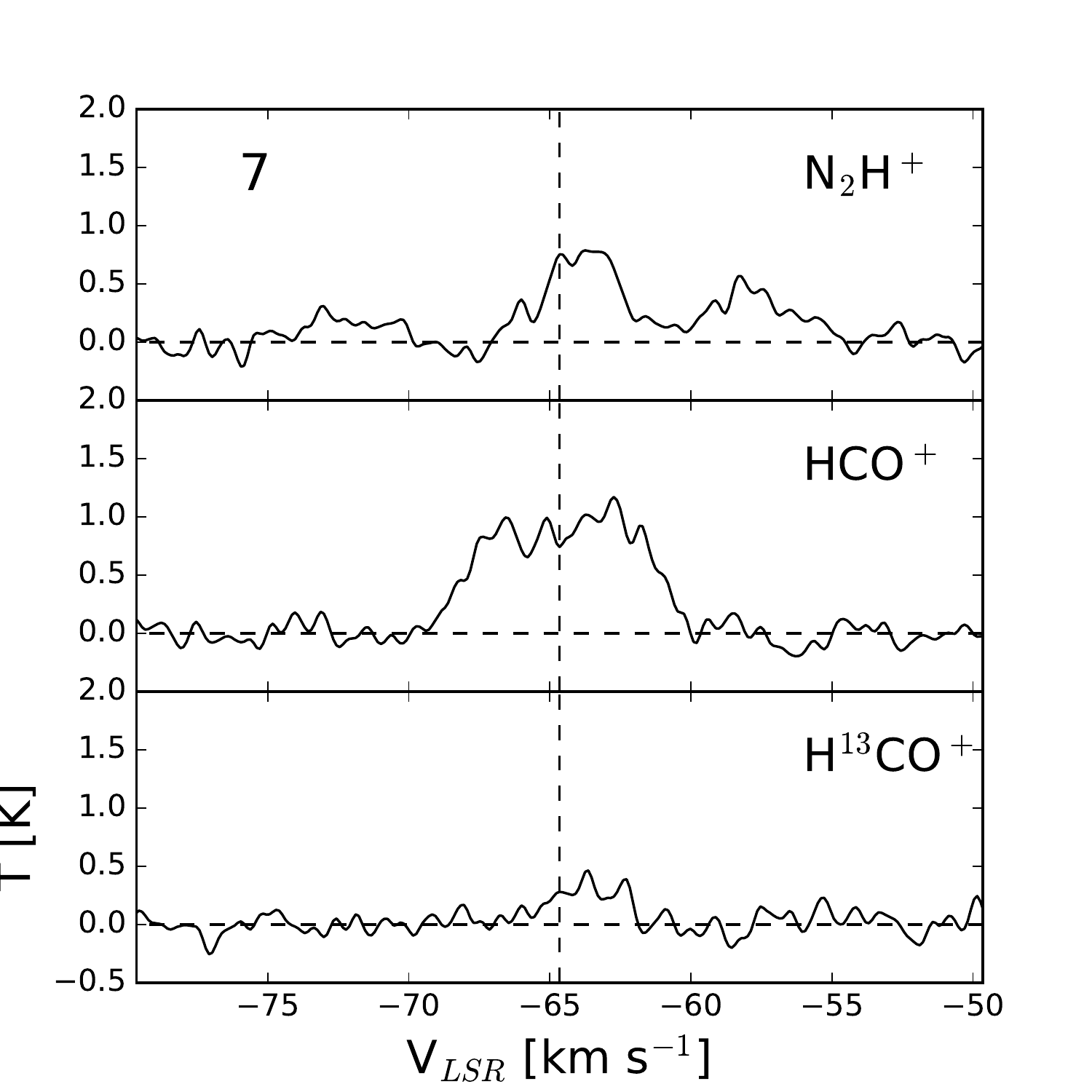}\hspace{-0.3cm}
\includegraphics[width=0.2\textwidth]{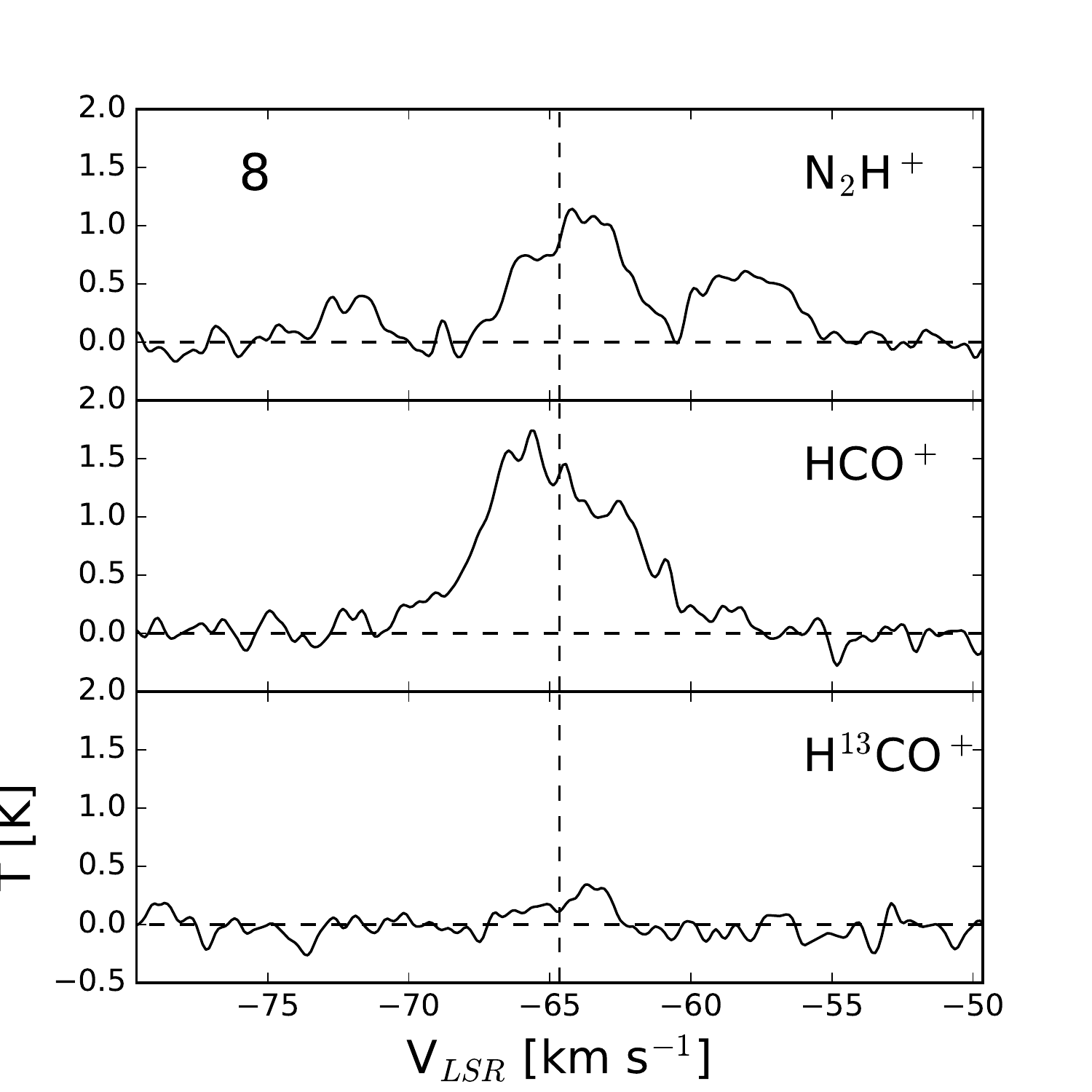}\hspace{-0.3cm}\\
\includegraphics[width=0.2\textwidth]{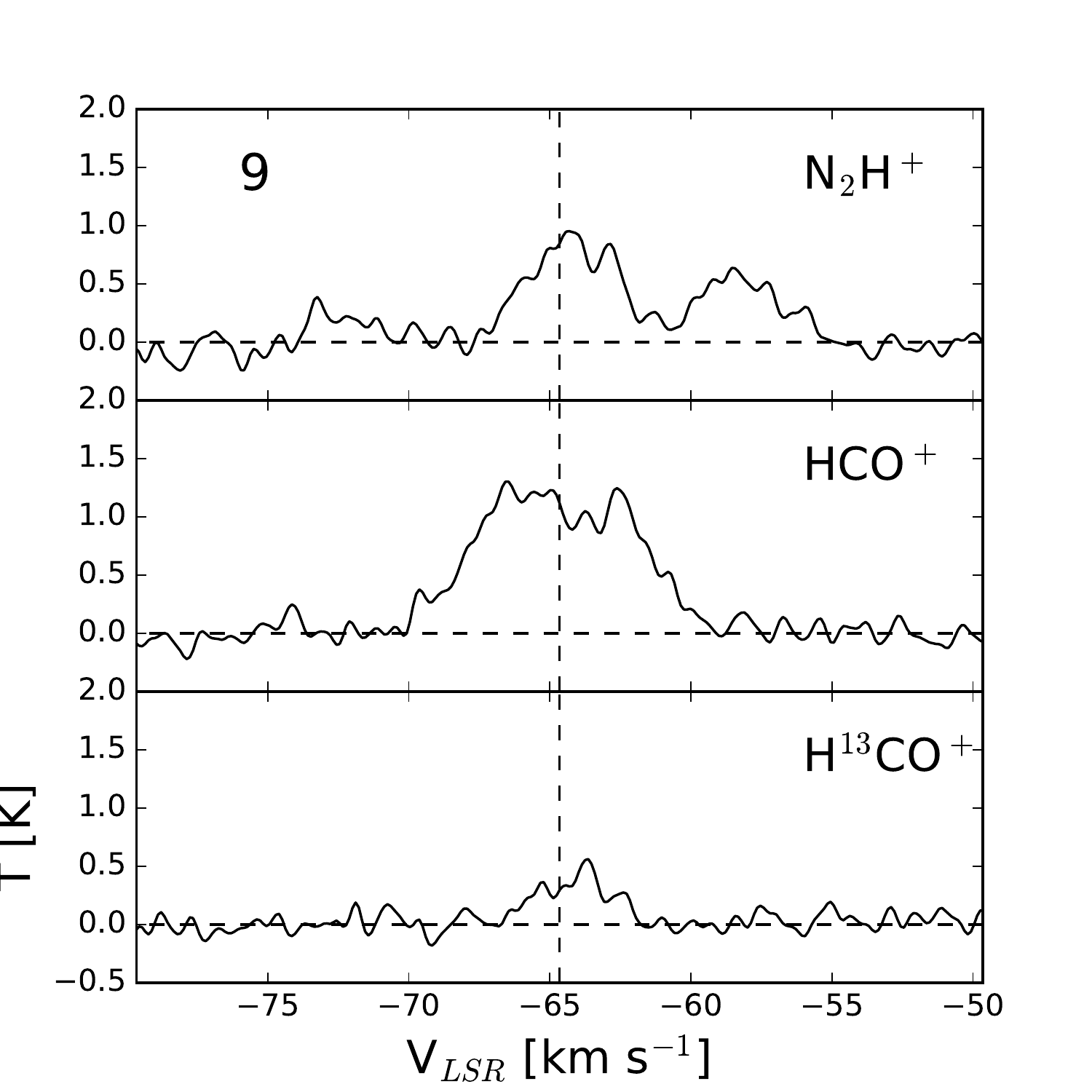}\hspace{-0.3cm}
\includegraphics[width=0.2\textwidth]{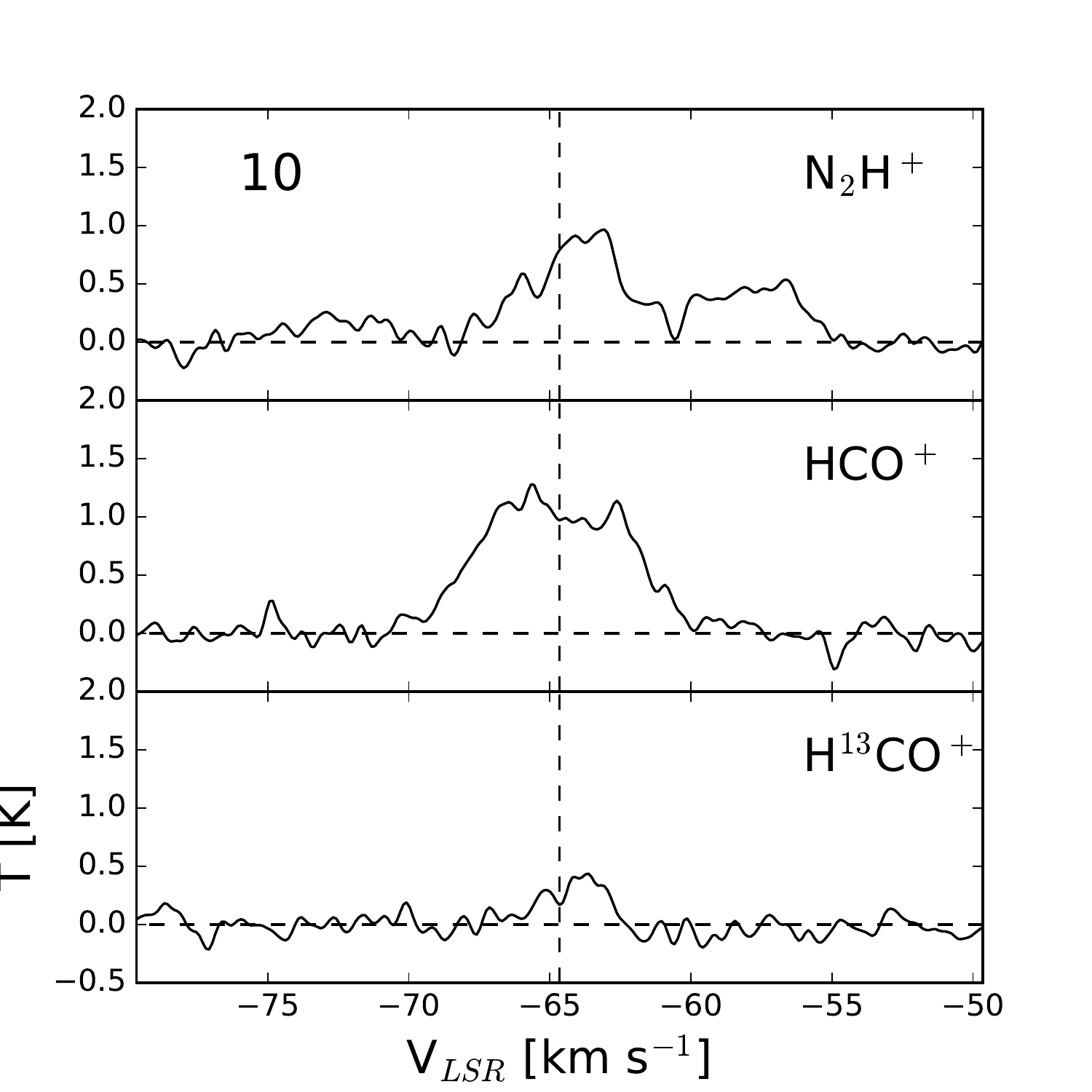}
\caption{Dust continuum and molecular line emission toward the YMC progenitor candidate \ymca. Upper left panel: ATLASGAL dust continuum emission overlaid with column density contours. Inner contours mark 70\% to 90\% of the column density peak in this clump and the outer contour marks a column density of 10$^{22}$ cm$^{-2}$. Upper right panel: Temperature of the clump overlaid with column density contours (values are the same as before). The levels of the column density are common to both dust peaks, indicating that it is single coherent structure. The dust temperature map shows a low dust temperature across 
the clump, suggesting an early stage of evolution. The lower panels show spectra  of N$_2$H$^+$, HCO$^+$, and H$^{13}$CO$^+$ at the positions indicated in the images (red numbers). The positions are chosen to illustrate the molecular emission around and at the position of the 2 dust continuum peaks seen toward this clump. 
 The HCO$^+$ spectrum of the clump show a blue peaked asymmetry in the whole clump, suggesting that it might be collapsing.}
 \label{ymc-1}
\end{figure*}

The column density map shows a common envelope that surrounds both dust continuum peaks (\ymca~and \ymcb). The dust temperature map shows an almost homogeneous, cold dust temperature across the clump (Figure \ref{ymc-1}). MALT90 observations show widespread emission of high density tracers N$_2$H$^+$, HNC, and HCO$^+$ (1-0). The morphology exhibited by these molecular lines is very similar, surrounding both dust continuum peaks, confirming that this is a physically coherent structure. 

The HCO$^{+}$ molecular line emission shows a blue peaked profile, which might indicate that this clump is collapsing. The blue peaked profile is confirmed by the emission of the optically thin tracer H$^{13}$CO$^{+}$ that has a peak at the velocity of the dip seen in HCO$^{+}$ (Figure \ref{ymc-1}). 

\begin{figure*}
\begin{center}
\includegraphics[trim=1cm 3.5cm 1cm 3.cm,clip,width=0.4\textwidth, angle=-90]{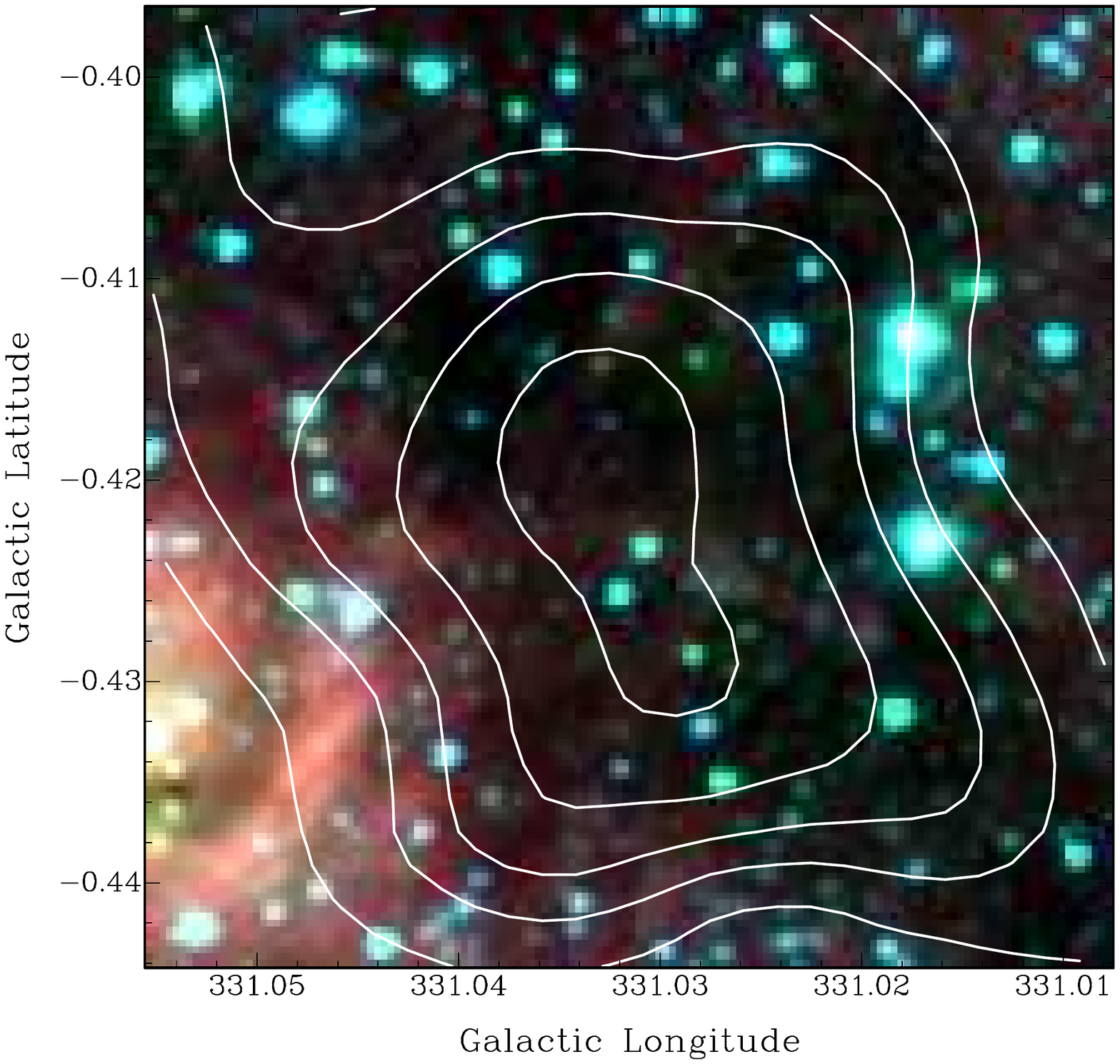}
\includegraphics[trim=1cm 3.5cm 1cm 3.cm,clip,width=0.4\textwidth, angle=-90]{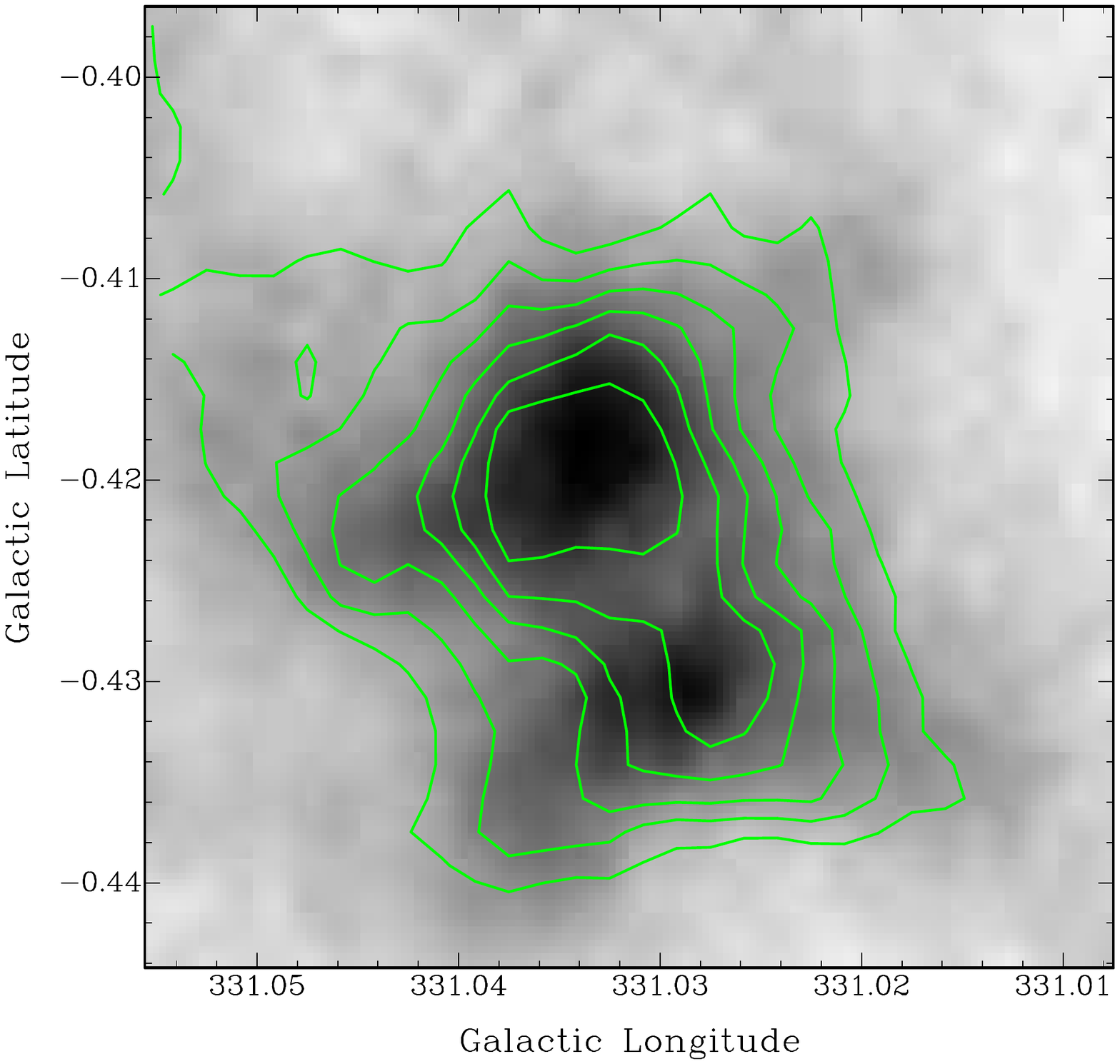}
\includegraphics[trim=1cm 3.5cm 1cm 3.cm,clip,width=0.4\textwidth, angle=-90]{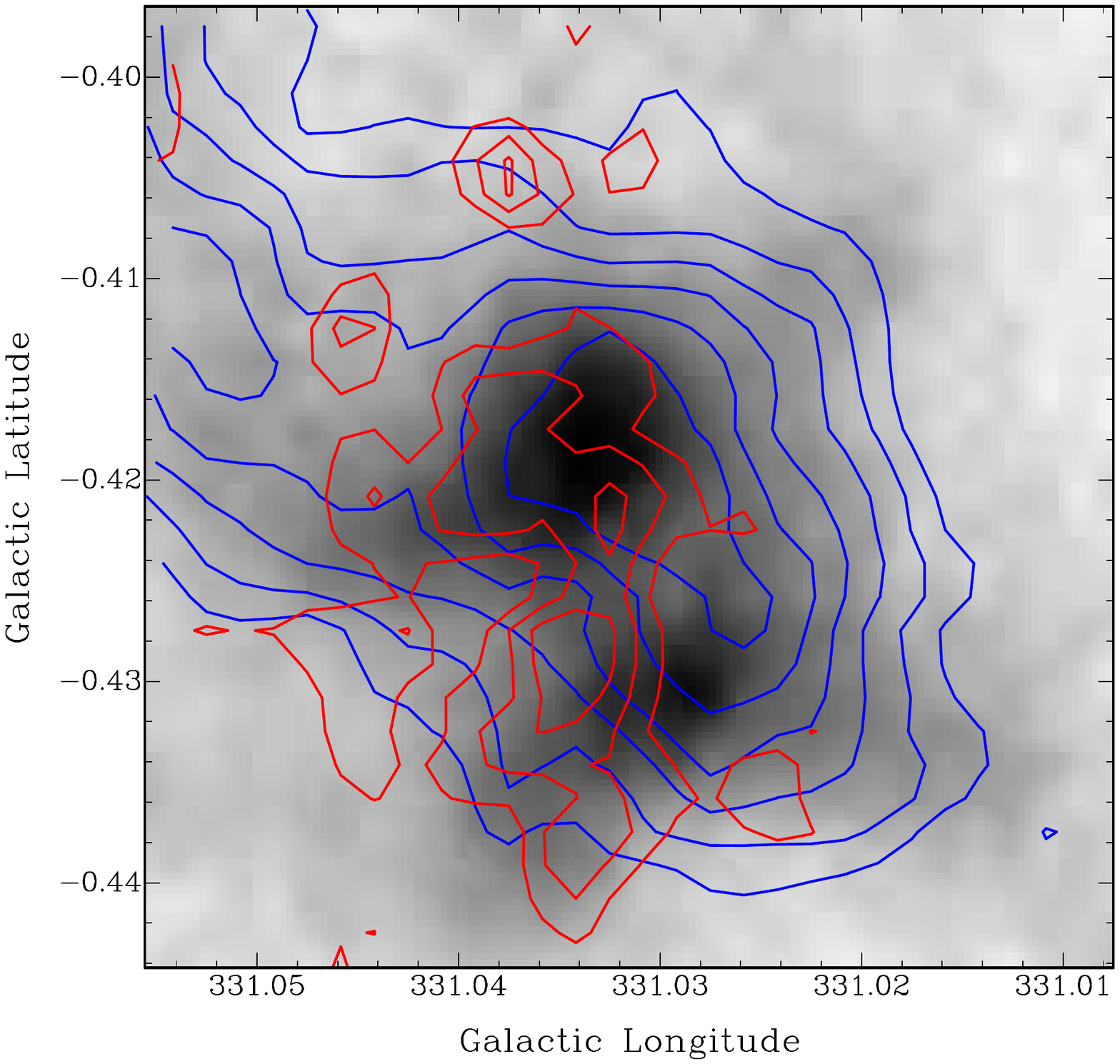}
\includegraphics[trim=1cm 3.5cm 1cm 3.cm,clip,width=0.4\textwidth, angle=-90]{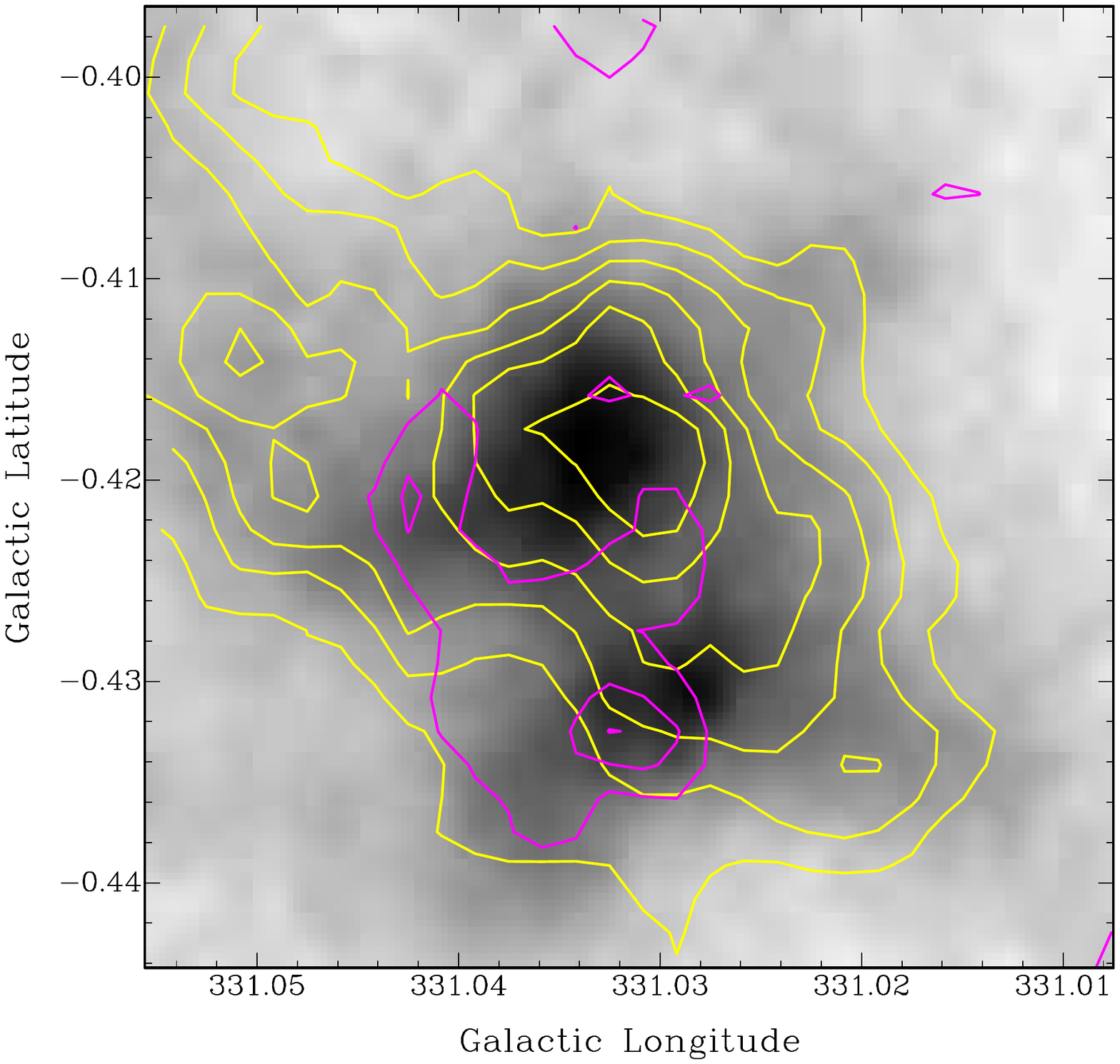}
\end{center}
\caption{Infrared, dust continuum and MALT90 molecular line emission toward \ymca\label{ymc-fig1}. Upper left panel: Infrared color image (red 8 \mum, green 4.5 \mum, and blue 3.6 \mum), overlaid with the column density determined toward this clump (as in Figure 1). This cloud does not appear as an infrared dark cloud, which is in agreement with the far kinematic distance determined via its molecular line emission. Upper right panel: N$_2$H$^+$ contours of the 40\% to 90\% peak emission overlaid on the ATLASGAL dust continuum emission map. Lower left panel: HCO$^+$ contours of the 40\% to 90\% peak emission in blue; H$^{13}$CO$^+$ contours of the 70\% to 90\% (5 $\sigma$ rms) of the peak emission in red. Lower right panel: HNC contours of the 40\% to 90\% peak emission in yellow; HN$^{13}$C contours of the 60\% to 90\% (5 $\sigma$ rms) of the peak emission in magenta. This YMC candidate shows coherent emission in the molecular lines, while in the continuum two over-densities can be identified which correspond to the clumps identified in the ATLASGAL catalog. The morphology is similar for N$_2$H$^+$, HNC, and HCO$^+$, with the peak of the molecular line emission located toward the upper dust over-density seen in the ATLASGAL maps. H$^{13}$CO$^+$ and HN$^{13}$C show different morphologies compared to the rest of the lines, with the peak of the emission between and closer to the lower dust over-density respectively. The overall morphology of the molecular emission exhibited by this clump, and its coherence in the (l,b,v) space, suggests that this clump has the mass and density to form a YMC.}\label{ymc-2}
\end{figure*}

The morphology exhibited by the optically thin tracers H$^{13}$CO$^+$ and HN$^{13}$C (1-0) does not follow what is seen in the other molecules, with the peak of the integrated intensity located between the 2 over-densities seen in the dust continuum maps (Figure \ref{ymc-2}). There is no emission from ``hot core chemistry molecules", such as HC$^{13}$CCN, HNCO, CH$_3$CN or HC$_3$N (1-0) that usually trace hot, dense gas that is heated by proto-stars, confirming the early stage of evolution of this clump. There is a marginal detection of the shock tracer SiO, which might indicate the presence of shocked gas, perhaps due to colliding gas inside this clump.

The apparent lack of any sign of current star formation, its mass and density, its coherence in the (l,b,v) space, and the blue-peaked profile indicative of infall motions makes \ymca~a excellent candidate of a YMC progenitor. Since the only YMC progenitor found is composed of two column density peaks, might argue that YMC progenitors are formed via the merging of two or more molecular clumps. We have to note that although our analysis places this clump at the far kinematic distance, further observations are needed to more accurately determine its distance and thus its true potential for forming a YMC. Nevertheless, the fact that we only see one, or possibly no, candidate for a YMC progenitor in this sample suggests that indeed YMCs are rare, or that the early stages of YMC formation are very quick. For example \citet{Svoboda-2016} estimate a life time of 0.03 Myr for YMC progenitors.

\subsection{Volume limited sample}

We analyzed the completeness of the MALT90 catalog as a function of Galactic longitude to determine a volume limited sample (VLS). The region where MALT90 is most complete, on which all the sources with fluxes $>0.25$ Jy from the ATLASGAL catalog were observed, lies between -50 and -24 degrees in Galactic longitude. Within this region the MALT90 survey reached a sensitivity of 200 M$_\odot$ up to 5 kpc. There are 670 clumps in this volume, which corresponds to nearly half of the clumps used in this study. To check that we did not have any bias due to the incompleteness of the survey in certain regions of the Galactic plane, we reanalyzed the data using only the VLS. We confirmed all  trends of the VLS followed the same trends exhibited by the whole sample. Therefore, considering that the inclusion of sources from less complete regions of the survey do not seem to introduce important biases, we keep the full sample to have better statistics in the analysis.

\section{Conclusions}

We determined the physical properties (mass, virial mass, virial state, volume density and bolometric luminosity) of \nmaltused~clumps located across the Galaxy.  Analyzing their molecular line emission and their dust column density, we determined that some column density peaks identified from the ATLASGAL maps are actually part of more extended coherent clumps.

Using the dust temperature of the clumps as proxy of their evolutionary stage, we determined how the physical properties of the clumps change as the clumps evolve. We found that the median mass of the clumps is $\sim10^3$ M$_\odot$ and shows no significant change with clump evolution. This suggests that we are sampling the same mass distribution in all evolutionary stages, and thus, that there is no mass bias in our sample. 

The volume density monotonically decreases with dust temperature, while the virial parameter and bolometric luminosity monotonically increases with dust temperature. The changes in these parameters are consistent with the evolution of the clumps. At the initial stages the clumps are colder ($\rm{T}_{dust}<20$ K), they are more bound ($\alpha\sim1$), their bolometric luminosity is lower ($\sim400$ L$_\odot$),  and their median volume density is higher ($\sim10^4$ cm$^{-3}$). This median volume density for the clumps in an early evolutionary stage is similar to the average star formation volume density of $10^4$ cm$^{-3}$ \citep{lada-2010}, suggesting that these clumps are dense enough to form stars.  As the clumps evolve and start forming stars their natal gas becomes hotter due to the radiation field of the newly forming stars, increasing their dust temperature ($>25 $ K) and increasing their bolometric luminosity ($\sim 2\times 10^4$ L$_\odot$). At the same time, the clumps start to dissipate their surrounding gas, becoming more gravitationally unbound ($\alpha>1$) and decreasing their mean volume density ($\sim5\times10^3$ cm$^{-3}$). This result suggests that dust temperature separation seems to be a better and more consistent indicator of the evolutionary stage of the clumps than classification based on their IR emission.

We also determined the potential of the clumps to evolve into stellar clusters harboring high-mass stars or to form open clusters and YMCs. Most the clumps satisfy the criteria for high-mass proto-cluster candidates. Their range of dust temperatures suggest that they are in a wide range of evolutionary stages. We identified 5 clumps with characteristics of YMC candidates, which may form stellar clusters with masses $\geq 10^4$ M$_\odot$. The small number of YMCs in our sample confirms that YMC progenitors in the Galactic plane are rare.

A significant fraction of the clumps (42\%) have low dust temperatures ($<$20 K) and therefore are good candidates for cluster-forming clumps in their earliest stage of evolution.  Most of these cold clumps have physical properties consistent with their being progenitors of clusters that will harbor high-mass stars. A sub-sample of these clumps appear dark at 70 \mum~\citep[far-IRDC clumps,][]{Guzman-2015}. These clumps have very cold median dust temperatures (14 K), high masses ($\sim10^4$ M$_\odot$), and high volume densities ($\sim10^4$ cm$^{-3}$); thus they likely represent the very earliest stages of high-mass star formation, and represent excellent targets to determine the initial conditions of high-mass star and cluster formation. 

Only one clump has the physical characteristics of a YMC progenitor in a very early stage of evolution. It has a low dust temperature (14 K), high mass and density and no sign of evident current star formation at infrared. So far, the only YMC progenitors known have been located in the harsh environment that surrounds the Galactic centre \citep[``The Brick'',][]{Longmore-2014, Rathborne-2014a, Rathborne-2014b, Rathborne-2015}. The fact that we have only found one clump that can potentially evolve into a YMC reinforces the suggestion that YMC progenitors are rare, and that their earliest stages of formation are very short.

\section*{Acknowledgements}
We thank the anonymous referee for all the valuable comments that have helped to improve this paper. Herschel is an ESA space observatory with science instruments provided by European-led Principal Investigator consortia and with important participation from NASA. A.E.G. acknowledges support from FONDECYT 3150570.
The MALT90 project team gratefully acknowledges the use
of dense clump positions supplied by ATLASGAL. ATLASGAL is a collaboration between the Max Planck Gesellschaft
(MPG: Max Planck Institute for Radioastronomy, Bonn and
the Max Planck Institute for Astronomy, Heidelberg), the
European Southern Observatory (ESO) and the University
of Chile.

\bibliography{bibliografia}

\newpage

\input{tabla_catalog}
\newpage

\input{tabla_resumen_new}

\input{tabla_ks}

\newpage

\appendix
\section{Derivation of the bolometric luminosity}

\begin{figure}
\includegraphics[width=0.4\textwidth]{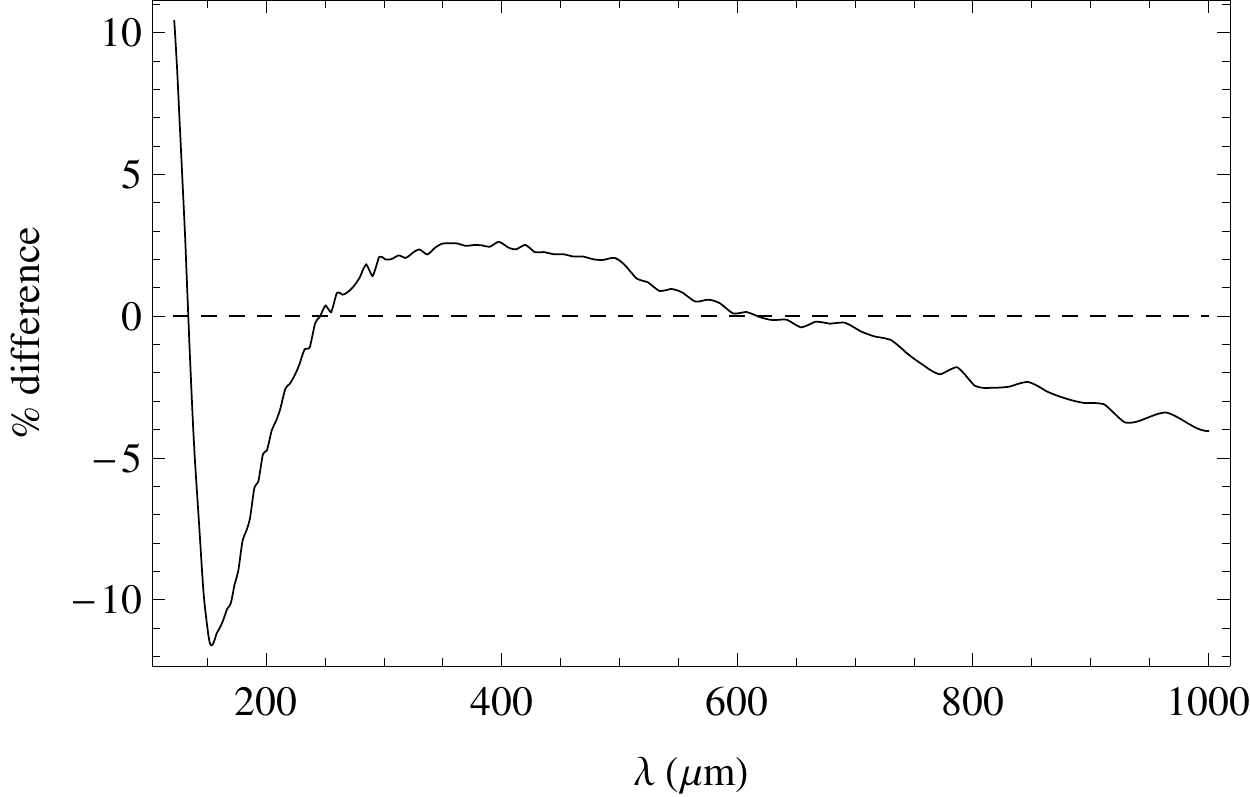}
\caption{Relative difference between Equation \eqref{eq-app} and the dust absorption 
curve used in Guzman et al.\ (2015)\label{fig-comp}}
\end{figure}

The bolometric luminosity of the clumps is derived from extrapolating the emission seen
in the continuum to shorter wavelengths. To do this we assume that the dust absorption
coefficient behaves as $\nu^\beta$ with frequency, where $\beta$ is the dust opacity spectral index. In the wavelength range
used to derive the dust temperature and column density by \citet{Guzman-2015}, the following is a good approximation for the mass absorption coefficient, assuming a gas-to-dust mass ratio of 100: 
\begin{equation}
\kappa_\nu=0.14\left(\frac{250~\mu{\rm m}}{\lambda_{\mu{\rm m}}}\right)^{\beta}~~\text{cm$^2$~g$^{-1}$},
\label{eq-app}
\end{equation}
\noindent where $\lambda$ is the frequency of the emission.
In Equation \eqref{eq-app}, $\beta=1.7$ gives a good fit. In Figure \ref{fig-comp} we show the percentage of difference between the approximation and the dust absorption coefficients from \citet{Ormel2011AA} used in \citet{Guzman-2015}. Note that the approximation worsens toward wavelengths shorter than 120 $\mu$m, thus Equation \eqref{eq-app} is only valid for dust temperatures lower than $\sim25$ K. For higher dust temperatures, a large fraction of the radiation is emitted by dust associated with an absorption curve significantly different than that given by Equation \eqref{eq-app}. 

The intensity received from a source of isothermal dust at temperature $T$ is given by $B_\nu(T)\left(1-e^{\tau_\nu}\right)$ where $\tau_\nu=\tau_0(\nu/\nu_0)^\beta$ is the optical depth. The bolometric intensity is given by
\begin{align}
I_{\rm bol}(T)&=\int_0^\infty B_\nu(T)(1-e^{\tau_\nu}) d\nu\label{eq-a2}\\
&=\frac{2k^4}{c^2 h^3}T^4\int_0^\infty \frac{x^3}{e^x-1}\left(1-\exp (\tau_T x^\beta)\right)dx\label{eq-a3}\\
&=\frac{15}{\pi^5}\sigma_{\rm SB}T^4\int_0^\infty \frac{x^3}{e^x-1}\left(1-\exp (\tau_T x^\beta)\right)dx~~,\label{eq-ibol}
\end{align}
where in \eqref{eq-a3} we make the substitution $x=h\nu/kT$. In Equations \eqref{eq-a3} and \eqref{eq-ibol}, $\tau_T$ is the opacity at the frequency $kT/h$ and $\sigma_{\rm SB}$ is the Stefan-Boltzmann's constant.

\begin{figure}
\includegraphics[width=0.4\textwidth]{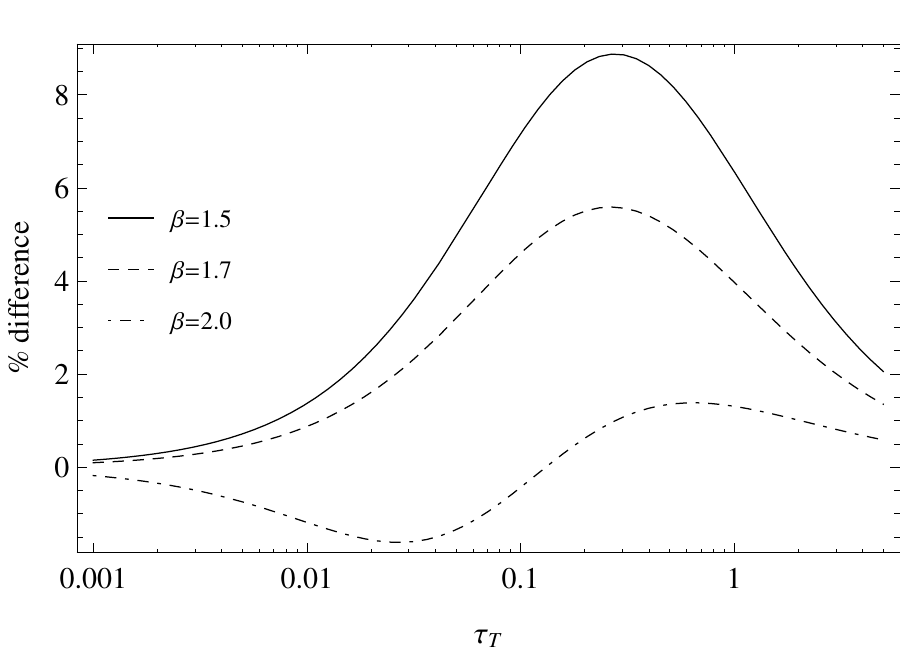}
\caption{Percentage of difference between the integral in \eqref{eq-ibol}
  and the approximation in Equation \eqref{eq-intapp} or three values of $\beta$.  \label{fig-num}}
\end{figure}

Asymptotic forms of the integral in the right hand side of Equation \eqref{eq-ibol} for small and large $\tau_T$  --- optically thin and thick cases, respectively --- are given by \citep[e.g.,][Chapter  6]{copson1965}
\begin{equation}
\begin{cases}
 \tau_T \Gamma(\beta+4)\zeta(\beta+4) +O(\tau_T^2) & \tau_T\ll1 \\
\frac{\pi^4}{15}+O(\tau_T^{-3/\beta}) & \tau_T\gg 1
\end{cases}~~,\label{eq-asym}
\end{equation}
where $\Gamma$ and $\zeta$ are the Gamma and Zeta functions, respectively and $O$  is the big-O notation that indicates the asymptotic behaviour in the approximation's residuals. In the optically thin case, the intensity (to first order) is proportional to $\tau_T$. In the optically thick case, on the other hand, the intensity reach an asymptotic constant value. To interpolate between these two regimes, we  use the following function
\begin{equation} 
\tau_T \Gamma(\beta+4)\zeta(\beta+4)\left(1+\frac{\tau_T \Gamma(\beta+4)\zeta(\beta+4)}{\pi^4/15}\right)^{-1}~~.\label{eq-intapp}
\end{equation}
Equation \eqref{eq-intapp}  has the same behaviour in the limit for small and large $\tau_T$ as Equation \eqref{eq-asym}, and therefore, as the integral in Equation \eqref{eq-ibol}. 
Figure \ref{fig-num} shows the difference between the exact integral and the approximation for different values of $\tau_T$ and $\beta$.

Finally, the bolometric luminosity of an isotropic source covering a solid angle $\Omega_s$, located at a distance $D$, is given by $L_{\rm bol}=I_{\rm bol}\Omega_sD^2$. Using the gas absorption coefficient $\kappa_\nu$ given in 
Equation \eqref{eq-app}  (which already assumes a gas-to-dust ratio of 100) we can write the opacity $\tau_T$ as 
$N_{\rm gas}\kappa_\nu$
with $\nu=kT/h$. 
Combining these expressions with Equations \eqref{eq-ibol} and \eqref{eq-intapp}, and changing to more convenient units, we write $L_{\rm bol}$ as:
\begin{align}
L_{\rm bol}&=40\,L_\odot \left(\frac{T}{\rm 10~K}\right)^{5.7} \left(\frac{D}{\rm kpc}\right)^{2} 
\left(\frac{\Omega_s}{\rm arcmin^2}\right)
\end{align}
\begin{align*}
 \times \left(\frac{N_{gas}}{\rm gr~cm^{-2}}\right)\times C(T,N_{\rm gas})~~,
\end{align*}
where $C(T,N_{\rm gas})$ is the opacity and dust temperature correction factor (the rightmost term in Equation \eqref{eq-intapp})
\begin{equation}
C(T,N_{\rm gas})=\left(1+1.6\times10^{-3} N_{\rm gas} T^{1.7}\right)^{-1}~~.
\end{equation}

\end{document}

%% file: tabla_catalog.tex
\begin{landscape}
\begin{table}
\caption{Example of the summary table of physical properties derived for the MALT90 clumps with reliable kinematic distances. The full table will be in electronic form. Column 1 shows the MALT90 name of the clumps. Clumps that are part of coherent groups are marked with the $\dagger$ symbol, and in the ``Notes" column is given the information of the associated MALT90 clump. Column 2 shows the kinematic distances (Whitaker et al. submitted). Column 3 corresponds to the consensus velocity. Column 4 correspons to the mean temperature of the clumps. Column 5 shows the area in parsec. Column 6 shows the effective deconvoluted radius. Column 7 shows the mass derived from the dust continuum emission. Column 8 shows the virial mass. Column 9 shows the virial parameter. Column 10 correspons to the volume density. Column 11 shows the bolometric luminosity derived from the extrapolation of the dust continuum emission and Column 12 shows the clumps associated in l-b-v space.}\label{tabla-catalogo}
\begin{tabular}{lrrrrrrrrrrc}
\hline
\hline
\multicolumn{1}{|c|}{Name}                &   \multicolumn{1}{|c|}{Dist.} & \multicolumn{1}{|c|}{$\Delta v$}   & \multicolumn{1}{|c|}{Temp.} & \multicolumn{1}{|c|}{Area}       & \multicolumn{1}{|c|}{R$_{eff}$} & \multicolumn{1}{|c|}{Mass}              & \multicolumn{1}{|c|}{Virial Mass}        & \multicolumn{1}{|c|}{$\alpha$} & \multicolumn{1}{|c|}{$n(H_2)$}        & \multicolumn{1}{|c|}{L$_{bol}$ }       & Notes \\
                     &   \multicolumn{1}{|c|}{(kpc)}         & \multicolumn{1}{|c|}{(km s$^{-1}$)}&  \multicolumn{1}{|c|}{(K) }       & \multicolumn{1}{|c|}{(pc$^2$)}   & \multicolumn{1}{|c|}{(pc)}      & \multicolumn{1}{|c|}{(10$^3$ M$_\odot$)}& \multicolumn{1}{|c|}{(10$^3$ M$_\odot$)} &          &  \multicolumn{1}{|c|}{(10$^4$ cm$^{-3}$)}   & \multicolumn{1}{|c|}{(10$^2$ L$_\odot$)}     &       \\
\hline\\
AGAL305.226$+$00.274\_S$^\dagger$ & 4.9 & 1.7$\pm$0.0 &  20$\pm$  2 & 6.4$\pm$1.7 & 1.4$\pm$0.2 & 7.2$\pm$2.0 &4.6$\pm$0.6 &0.6$\pm$0.2 &1.5$\pm$0.2 &390.8$\pm$117.0 & AGAL305.237+00.286\_S \\ 
AGAL305.234$+$00.262\_S & 5.3 & 1.5$\pm$0.0 &  23$\pm$  2 & 0.0$\pm$0.0 & 0.1$\pm$0.0 & 2.5$\pm$0.9 &0.2$\pm$0.0 &0.1$\pm$0.0 &1133.4$\pm$205.7 &4.1$\pm$1.7 &  \\ 
AGAL305.236$-$00.022\_S & 7.1 & 1.6$\pm$0.0 &  18$\pm$  2 & 14.1$\pm$3.5 & 2.1$\pm$0.3 & 10.3$\pm$2.6 &4.2$\pm$0.6 &0.4$\pm$0.1 &0.4$\pm$0.1 &428.4$\pm$136.1 &  \\ 
AGAL305.242$-$00.041\_S & 6.8 & 1.6$\pm$0.1 &  16$\pm$  2 & 0.8$\pm$0.3 & 0.5$\pm$0.1 & 3.9$\pm$1.2 &1.1$\pm$0.2 &0.3$\pm$0.1 &11.6$\pm$1.9 &10.8$\pm$4.0 &  \\ 
AGAL305.242$+$00.276\_S & 4.9 & 1.4$\pm$0.0 &  22$\pm$  6 & 1.5$\pm$0.6 & 0.7$\pm$0.1 & 0.8$\pm$0.3 &1.1$\pm$0.2 &1.3$\pm$0.5 &1.0$\pm$0.2 &309.2$\pm$131.0 &  \\ 
AGAL305.257$+$00.004\_S & 6.8 & 1.4$\pm$0.2 &  37$\pm$  2 & 1.5$\pm$0.5 & 0.7$\pm$0.1 & 0.2$\pm$0.1 &1.2$\pm$0.3 &7.0$\pm$3.0 &0.2$\pm$0.0 &286.2$\pm$105.6 &  \\ 
AGAL305.259$+$00.326\_S & 5.2 & 1.6$\pm$0.0 &  18$\pm$  2 & 0.2$\pm$0.1 & 0.2$\pm$0.0 & 1.3$\pm$0.5 &0.5$\pm$0.1 &0.4$\pm$0.2 &37.9$\pm$6.9 &2.3$\pm$1.0 &  \\ 
AGAL305.271$-$00.029\_S & 6.7 & 2.0$\pm$0.1 &  22$\pm$  1 & 2.4$\pm$1.0 & 0.9$\pm$0.2 & 1.0$\pm$0.4 &2.9$\pm$0.6 &2.8$\pm$1.3 &0.6$\pm$0.1 &65.2$\pm$28.7 &  \\ 
AGAL305.271$-$00.009\_S & 6.7 & 1.9$\pm$0.0 &  27$\pm$  2 & 12.0$\pm$4.7 & 2.0$\pm$0.4 & 9.3$\pm$3.6 &5.7$\pm$1.1 &0.6$\pm$0.3 &0.5$\pm$0.1 &1687.7$\pm$738.1 &  \\ 
AGAL305.272$+$00.296\_S & 5.6 & 1.2$\pm$0.1 &  26$\pm$  1 & 3.6$\pm$1.3 & 1.1$\pm$0.2 & 2.7$\pm$1.0 &1.2$\pm$0.2 &0.4$\pm$0.2 &0.9$\pm$0.2 &258.9$\pm$104.6 &  \\ 
AGAL305.346$+$00.212\_S & 4.9 & 2.7$\pm$0.1 &  32$\pm$  1 & 2.2$\pm$0.8 & 0.8$\pm$0.1 & 2.6$\pm$0.9 &4.9$\pm$0.9 &1.8$\pm$0.7 &1.8$\pm$0.3 &959.3$\pm$384.7 &  \\ 
AGAL305.357$+$00.202\_S & 5.8 & 2.3$\pm$0.1 &  32$\pm$  1 & 3.1$\pm$1.1 & 1.0$\pm$0.2 & 3.6$\pm$1.3 &4.4$\pm$0.8 &1.2$\pm$0.5 &1.5$\pm$0.3 &1315.4$\pm$531.6 &  \\ 
AGAL305.361$+$00.186\_S & 5.5 & 1.4$\pm$0.0 &  26$\pm$  1 & 4.3$\pm$1.5 & 1.2$\pm$0.2 & 5.5$\pm$1.9 &1.8$\pm$0.3 &0.3$\pm$0.1 &1.4$\pm$0.3 &1887.1$\pm$758.1 &  \\ 
AGAL305.367$+$00.212\_S & 6.5 & 1.8$\pm$0.1 &  28$\pm$  1 & 5.9$\pm$2.2 & 1.4$\pm$0.3 & 5.6$\pm$2.1 &3.6$\pm$0.8 &0.6$\pm$0.3 &0.9$\pm$0.2 &1402.3$\pm$598.2 &  \\ 
AGAL305.381$+$00.241\_S & 5.6 & 1.9$\pm$0.3 &  23$\pm$  4 & 1.5$\pm$0.5 & 0.7$\pm$0.1 & 1.1$\pm$0.4 &2.0$\pm$0.8 &1.8$\pm$0.9 &1.4$\pm$0.3 &71.3$\pm$28.7 &  \\ 
AGAL305.509$+$00.367\_S & 5.9 & 1.7$\pm$0.3 &  23$\pm$  2 & 1.7$\pm$0.6 & 0.7$\pm$0.1 & 0.4$\pm$0.1 &1.7$\pm$0.8 &4.1$\pm$2.3 &0.4$\pm$0.1 &51.7$\pm$21.0 &  \\ 
AGAL305.512$+$00.334\_S & 4.9 & 1.1$\pm$0.2 &  21$\pm$  1 & 1.0$\pm$0.4 & 0.6$\pm$0.1 & 0.2$\pm$0.1 &0.5$\pm$0.2 &2.4$\pm$1.3 &0.5$\pm$0.1 &18.4$\pm$7.7 &  \\ 
AGAL305.512$+$00.349\_S & 5.6 & 1.4$\pm$0.2 &  24$\pm$  2 & 2.0$\pm$0.7 & 0.8$\pm$0.1 & 0.5$\pm$0.2 &1.2$\pm$0.4 &2.5$\pm$1.3 &0.4$\pm$0.1 &79.0$\pm$31.7 &  \\ 
AGAL305.521$+$00.361\_S & 6.0 & 2.4$\pm$0.3 &  27$\pm$  3 & 1.7$\pm$0.6 & 0.7$\pm$0.1 & 0.3$\pm$0.1 &3.4$\pm$1.1 &11.6$\pm$5.7 &0.3$\pm$0.1 &87.0$\pm$35.5 &  \\ 
AGAL305.536$-$00.019\_S & 5.4 & 2.4$\pm$0.3 &  33$\pm$  3 & 1.1$\pm$0.4 & 0.6$\pm$0.1 & 0.2$\pm$0.1 &2.7$\pm$0.8 &13.8$\pm$6.3 &0.4$\pm$0.1 &151.1$\pm$60.6 &  \\ 
AGAL305.539$+$00.339\_S & 6.5 & 1.3$\pm$0.1 &  24$\pm$  2 & 5.0$\pm$1.9 & 1.3$\pm$0.2 & 2.6$\pm$1.0 &1.7$\pm$0.4 &0.7$\pm$0.3 &0.5$\pm$0.1 &279.2$\pm$119.2 &  \\ 
AGAL305.549$+$00.004\_S & 5.9 & 1.7$\pm$0.2 &  33$\pm$  2 & 1.3$\pm$0.5 & 0.7$\pm$0.1 & 0.3$\pm$0.1 &1.6$\pm$0.4 &5.0$\pm$2.3 &0.5$\pm$0.1 &248.6$\pm$101.0 &  \\ 
AGAL305.554$-$00.011\_S & 3.8 & 1.6$\pm$0.3 &  26$\pm$  1 & 2.1$\pm$1.2 & 0.8$\pm$0.2 & 0.8$\pm$0.4 &1.7$\pm$0.7 &2.2$\pm$1.6 &0.6$\pm$0.2 &260.4$\pm$158.2 &  \\ 
AGAL305.562$+$00.014\_S & 5.4 & 1.2$\pm$0.1 &  32$\pm$  1 & 3.8$\pm$1.3 & 1.1$\pm$0.2 & 1.9$\pm$0.7 &1.3$\pm$0.3 &0.7$\pm$0.3 &0.6$\pm$0.1 &852.5$\pm$342.7 &  \\ 
AGAL305.574$-$00.342\_S & 4.9 & 1.5$\pm$0.3 &  21$\pm$  1 & 0.5$\pm$0.1 & 0.4$\pm$0.0 & 0.1$\pm$0.0 &0.8$\pm$0.3 &10.3$\pm$4.2 &0.4$\pm$0.0 &5.0$\pm$1.3 &  \\ 
AGAL305.581$+$00.037\_S & 6.0 & 2.0$\pm$0.3 &  34$\pm$  1 & 0.7$\pm$0.3 & 0.5$\pm$0.1 & 0.4$\pm$0.1 &1.6$\pm$0.5 &4.4$\pm$2.1 &1.4$\pm$0.3 &54.4$\pm$22.1 &  \\ 
AGAL305.589$+$00.462\_S & 3.8 & 1.3$\pm$0.2 &  16$\pm$  2 & 0.1$\pm$0.1 & 0.2$\pm$0.1 & 0.4$\pm$0.2 &0.3$\pm$0.1 &0.6$\pm$0.4 &22.7$\pm$6.6 &0.9$\pm$0.5 &  \\ 
AGAL305.777$-$00.249\_S & 2.7 & 1.0$\pm$0.1 &  19$\pm$  3 & 0.1$\pm$0.1 & 0.2$\pm$0.1 & 0.2$\pm$0.1 &0.2$\pm$0.1 &0.8$\pm$0.6 &14.9$\pm$4.8 &1.4$\pm$0.9 &  \\ 
AGAL305.794$-$00.096\_S & 5.0 & 1.4$\pm$0.1 &  16$\pm$  2 & 1.3$\pm$0.5 & 0.6$\pm$0.1 & 1.6$\pm$0.5 &1.1$\pm$0.2 &0.7$\pm$0.3 &2.4$\pm$0.4 &9.6$\pm$3.9 &  \\ 
\hline
\hline
\end{tabular}
\end{table}
\end{landscape}

%% file: tabla_resumen_new.tex
\begin{table*}
\begin{center}
\caption{Summary of physical properties derived for the full sample and divided by dust temperature of the clumps.}\label{tabla-resumen}
\begin{tabular}{crrrrrrrrrrrr}
\hline
\hline
Sample &\%& \multicolumn{4}{|c|}{Temperature} & \multicolumn{4}{|c|}{$R_{eff}$}  \\
 & &\multicolumn{4}{|c|}{K} & \multicolumn{4}{|c|}{pc}  \\
&&Median & $\sigma_{Med.}$ & Min & Max& Median & $\sigma_{Med.}$& Min & Max \\
\hline
\vspace{0.2cm}
Full Sample &&$ 22$&$5.9$ &$ 10$ &$ 44$ &$0.82$ &$0.6$&$0.03$ &$8.7$   \\
T$_{d}\leq15$ &12&$ 14$ &$1.5$ &$ 10$ &$ 15$ &$0.64$&$0.46$ &$0.071$ &$3.1$  \\
$15<$T$_{d}\leq20$ &30&$ 18$ &$1.6$ &$ 15$ &$ 20$&$0.74$ &$0.55$ &$0.033$ &$5.7$  \\
$20<$T$_{d}\leq25$ &30&$ 22$ &$1.9$ &$ 20$ &$ 25$&$0.83$ &$0.59$ &$0.03$ &$8.7$  \\
$25<$T$_{d}\leq30$ &19&$ 27$ &$1.5$ &$ 25$ &$ 30$&$0.93$ &$0.66$ &$0.061$ &$4.4$  \\
\vspace{0.2cm}
$30<$T$_{d}$       & 9&$ 33$ &$2.8$ &$ 30$ &$ 44$&$0.77$ &$0.54$ &$0.041$ &$3.9$  \\
HMPC               &78&$ 21$ &$5.9$ &$ 10$&$ 44$ &$0.92$ &$0.68$ &$0.03$ &$8.7$ \\
YMC                &0.4&$ 20$ &$  3$ &$ 14$&$ 23$ &$3.6$ &$1.6$ &$2.5$ &$5.7$  \\
\vspace{0.2cm}
OC   &67&$ 22$ &$5.9$ &$ 10$&$ 44$ &$0.97$ &$0.58$ &$0.4$ &$3.7$ \\
Cold, Dark 70 $\mu$m               & 2&$ 14$ &$2.3$&$ 10$ &$ 19$ &$0.7$ &$0.37$ &$0.061$ &$2.1$   \\
\hline
\hline
Sample && \multicolumn{4}{|c|}{Area} & \multicolumn{4}{|c|}{Mass}\\
 & &\multicolumn{4}{|c|}{pc$^2$} & \multicolumn{4}{|c|}{M$_\odot$} \\
&&Median & $\sigma_{Med.}$ & Min & Max& Median & $\sigma_{Med.}$& Min & Max \\
\hline
\vspace{0.2cm}
Full Sample &&$2.1$ &$2.5$&$0.0029$ &$2.4e+02$ &$1e+03$ &$1.2e+03$ &$0.99$ &$1e+05$ \\
T$_{d}\leq15$ &&$1.3$ &$1.5$&$0.016$ &$ 31$ &$9.8e+02$ &$1.2e+03$ &$ 17$ &$4.1e+04$ \\
$15<$T$_{d}\leq20$ &&$1.7$ &$2.2$&$0.0033$ &$1e+02$ &$1e+03$ &$1.1e+03$ &$ 12$ &$9.9e+04$ \\
$20<$T$_{d}\leq25$ &&$2.2$ &$2.5$&$0.0029$ &$2.4e+02$ &$1.1e+03$ &$1.3e+03$ &$2.3$ &$1e+05$ \\
$25<$T$_{d}\leq30$ &&$2.7$ &$3.2$&$0.012$ &$ 61$ &$1.2e+03$ &$1.4e+03$ &$ 18$ &$4.5e+04$ \\
\vspace{0.2cm}
$30<$T$_{d}$       &&$1.9$ &$2.1$&$0.0052$ &$ 47$ &$6.3e+02$ &$8e+02$ &$0.99$ &$1.6e+04$ \\
HMPC               &&$2.7$ &$3.3$&$0.0029$ &$2.4e+02$ &$1.5e+03$ &$1.6e+03$ &$ 12$ &$1e+05$ \\
YMC                 &&$ 41$ &$ 31$&$ 20$ &$1e+02$ &$6.7e+04$ &$4.8e+04$ &$3.3e+04$ &$1e+05$ \\
\vspace{0.2cm}
OC   &&$  3$ &$3.1$&$0.5$ &$ 42$ &$1e+03$ &$1.1e+03$ &$ 37$ &$1.4e+04$  \\
Cold, Dark 70 $\mu$m               &&$1.5$ &$1.3$&$0.012$ &$ 14$ &$1.2e+03$ &$1.2e+03$ &$1.9e+02$ &$2.5e+04$ \\
\hline
\hline
Sample &~~&\multicolumn{4}{|c|}{Virial Mass}& \multicolumn{4}{|c|}{$\alpha$} \\
 & & \multicolumn{4}{|c|}{ M$_\odot$} & \multicolumn{4}{|c|}{}\\
&&Median & $\sigma_{Med.}$ & Min & Max& Median & $\sigma_{Med.}$& Min & Max \\
\hline\vspace{0.2cm}
Full Sample &&$1.2e+03$ &$1.1e+03$ &$ 21$ &$3.8e+04$ &$1.1$ &$0.95$ &$0.036$ &$2.4e+02$ \\
T$_{d}\leq15$ &&$8.4e+02$ &$8.5e+02$ &$ 75$ &$1.3e+04$ &$0.8$ &$0.71$ &$0.036$ &$ 30$ \\
$15<$T$_{d}\leq20$ &&$9.9e+02$ &$9.4e+02$ &$ 21$ &$3.7e+04$ &$0.94$ &$0.86$ &$0.053$ &$ 20$ \\
$20<$T$_{d}\leq25$ &&$1.2e+03$ &$9.6e+02$ &$ 21$ &$3.4e+04$ &$0.96$ &$0.87$ &$0.071$ &$1.4e+02$ \\
$25<$T$_{d}\leq30$ &&$1.5e+03$ &$1.5e+03$ &$ 63$ &$3.8e+04$ &$1.3$ &$1.1$ &$0.11$ &$ 32$ \\
\vspace{0.2cm}
$30<$T$_{d}$ &&$1.6e+03$ &$1.6e+03$ &$ 75$ &$1.9e+04$ &$2.1$ &$  2$ &$0.19$ &$2.4e+02$ \\
HMPC &&$1.3e+03$ &$1.2e+03$ &$ 21$ &$3.8e+04$ &$0.82$ &$0.64$ &$0.036$ &$ 19$ \\
YMC& &$1.1e+04$ &$3.3e+03$ &$6.6e+03$ &$2.5e+04$ &$0.21$ &$0.067$ &$0.067$ &$0.36$ \\
\vspace{0.2cm}
OC& &$1.3e+03$ &$1.1e+03$ &$ 21$ &$3.8e+04$ &$1.3$ &$0.95$ &$0.11$ &$ 41$ \\
Cold, Dark 70 $\mu$m &&$1.1e+03$ &$1.1e+03$ &$ 51$ &$3.9e+03$ &$0.51$ &$0.43$ &$0.15$ &$7.5$ \\
\hline
\hline
Sample &~~& \multicolumn{4}{|c|}{$n(H_2)$}& \multicolumn{4}{|c|}{L$_{bol}$ }\\
 & & \multicolumn{4}{|c|}{gr cm$^{-3}$}& \multicolumn{4}{|c|}{L$_\odot$ }\\
&&Median & $\sigma_{Med.}$ & Min & Max& Median & $\sigma_{Med.}$& Min & Max \\
\hline
\vspace{0.2cm}
Full Sample &&$6.8e+03$ &$6.7e+03$ &$4.9e+02$ &$1.1e+07$ &$4.9e+03$ &$6.8e+03$ &$4.9$ &$1.4e+06$ \\
T$_{d}\leq15$ &&$1.2e+04$ &$1.1e+04$ &$1.6e+03$ &$1.2e+06$ &$4.3e+02$ &$5.3e+02$ &$6.3$ &$3.1e+04$ \\
$15<$T$_{d}\leq20$ &&$7.3e+03$ &$7e+03$ &$8.7e+02$ &$6e+06$ &$1.7e+03$ &$2.1e+03$ &$5.5$ &$5e+05$ \\
$20<$T$_{d}\leq25$ &&$6.2e+03$ &$6.1e+03$ &$4.9e+02$ &$1.1e+07$ &$7.3e+03$ &$8.8e+03$ &$4.9$ &$1.1e+06$ \\
$25<$T$_{d}\leq30$ &&$5.2e+03$ &$5.6e+03$ &$6.6e+02$ &$2.5e+06$ &$1.9e+04$ &$2.2e+04$ &$ 67$ &$6.9e+05$ \\
\vspace{0.2cm}
$30<$T$_{d}$ &&$4.5e+03$ &$4.3e+03$ &$7.5e+02$ &$6.1e+05$ &$3e+04$ &$3.8e+04$ &$ 62$ &$1.4e+06$ \\
HMPC &&$8.5e+03$ &$8.6e+03$ &$4.9e+02$ &$1.1e+07$ &$6.3e+03$ &$8.7e+03$ &$4.9$ &$1.4e+06$ \\
YMC& &$5.9e+03$ &$3.6e+03$ &$2.2e+03$ &$8.6e+03$ &$1.8e+05$ &$2.2e+05$ &$3.1e+04$ &$5.4e+05$ \\
\vspace{0.2cm}
OC& &$4.7e+03$ &$3.9e+03$ &$6.6e+02$ &$1.1e+05$ &$6.9e+03$ &$8.8e+03$ &$ 44$ &$1e+06$\\
Cold, Dark 70 $\mu$m &&$1e+04$ &$5.8e+03$ &$2.5e+03$ &$6e+06$ &$8.5e+02$ &$9.1e+02$ &$5.5$ &$6.9e+03$ \\
\hline
\hline
\end{tabular}
\end{center}
\end{table*}

%% file: tabla_ks.tex
\begin{table*}
\begin{center}
\caption{KS statistical test values for the distribution of mass, density, bolometric luminosity and virial parameter as function of dust temperature bins.}\label{kstest}
\begin{tabular}{ccccccccccccccccccccccccccc}
\hline
\hline
Sample & \multicolumn{2}{|c|}{Mass} & \multicolumn{2}{|c|}{$n(H_2)$}& \multicolumn{2}{|c|}{L$_{bol}$ }& \multicolumn{2}{|c|}{$\alpha$ }\\
& KS Value & p Value& KS Value & p Value& KS Value & p Value& KS Value & p Value \\
\hline\\
T$_{d}\leq15$ &      4.41e-02 &9.60e-01&2.49e-01 &1.49e-07&4.87e-01 &1.41e-27&1.40e-01 &1.15e-02\\
$15<$T$_{d}\leq20$ & 3.67e-02 &8.28e-01&5.63e-02 &3.15e-01&2.49e-01 &4.40e-16&5.78e-02 &2.84e-01\\
$20<$T$_{d}\leq25$ & 3.84e-02 &7.80e-01&3.50e-02 &8.66e-01&1.30e-01 &1.05e-04&5.11e-02 &4.29e-01\\
$25<$T$_{d}\leq30$ & 6.63e-02 &3.39e-01&1.13e-01 &1.18e-02&3.24e-01 &8.23e-19&7.98e-02 &1.54e-01\\
$30<$T$_{d}$ &       1.62e-01 &6.83e-03&1.52e-01 &1.29e-02&4.02e-01 &1.12e-15&2.55e-01 &1.46e-06\\
\hline
\hline
\end{tabular}
\end{center}
\end{table*}